 \documentclass[final,3p,times]{elsarticle}
\usepackage{amssymb}
\usepackage{multirow}
\usepackage{subcaption}
\usepackage{amsmath}
\usepackage{comment}
\usepackage{hyperref}
\usepackage{xcolor}
\begin{document}

\begin{frontmatter}

%% Title, authors and addresses

%% use the tnoteref command within \title for footnotes;
%% use the tnotetext command for theassociated footnote;
%% use the fnref command within \author or \address for footnotes;
%% use the fntext command for theassociated footnote;
%% use the corref command within \author for corresponding author footnotes;
%% use the cortext command for theassociated footnote;
%% use the ead command for the email address,
%% and the form \ead[url] for the home page:
%% \title{Title\tnoteref{label1}}
%% \tnotetext[label1]{}
%% \author{Name\corref{cor1}\fnref{label2}}
%% \ead{email address}
%% \ead[url]{home page}
%% \fntext[label2]{}
%% \cortext[cor1]{}
%% \affiliation{organization={},
%%             addressline={},
%%             city={},
%%             postcode={},
%%             state={},
%%             country={}}
%% \fntext[label3]{}

\title{Improving ASR Fairness for Cleft Lip and Palate Speech: A Study on Severity-Aware Data Mixing}%Improving ASR Fairness for Cleft Lip and Palate Speech: A Study on Severity-Aware Aug
    %Fairness of Automatic Speech Recognition in Cleft Lip and Palate Speech

%\author[1]{Susmita Bhattacharjee\corref{cor1}}
%\ead{sbhattacharjee@iitg.ac.in}

%\author[2]{Jagabandhu Mishra\corref{cor1}}
%\ead{jagabandhu.mishra@uef.fi}

%\author[1]{H.S. Shekhawat}
%\ead{h.s.shekhawat@iitg.ac.in}

%\author[3]{S. R. Mahadeva Prasanna}
%\ead{prasanna@iitdh.ac.in}

\author[1]{\texorpdfstring{Susmita Bhattacharjee\corref{cor1}}{Susmita Bhattacharjee}}
\ead{sbhattacharjee@iitg.ac.in}

\author[2]{\texorpdfstring{Jagabandhu Mishra\corref{cor1}}{Jagabandhu Mishra}}
\ead{jagabandhu.mishra@uef.fi}

\author[1]{H.S. Shekhawat}
\ead{h.s.shekhawat@iitg.ac.in}

\author[3,4]{Ravi Jasuja}
\ead{jasuja1@gmail.com}

\author[5]{S. R. Mahadeva Prasanna}
\ead{prasanna@iiitdwd.ac.in}

%\cortext[cor1]{Corresponding author.}
\cortext[cor1]{Corresponding authors: Susmita Bhattacharjee}

\affiliation[1]{organization={Department of Electronics and Electrical Engineering},
            addressline={Indian Institute of Technology Guwahati}, 
            city={Guwahati},
            postcode={781039}, 
            state={Assam},
            country={India}}

\affiliation[2]{organization={School of Computing},
            addressline={University of Eastern Finland}, 
            city={Joensuu},
            country={Finland}}         

\affiliation[3]{%
  organization={Brigham and Women's Hospital, Harvard Medical School},
  city={Boston},
  state={MA},
  country={United States}
}

\affiliation[4]{%
  organization={Function Promoting Therapies},
  city={Waltham},
  state={MA},
  country={United States}
}

\affiliation[5]{organization={Department of Electrical Engineering}, 
            addressline={Indian Institute of Information Technology Dharwad (IIIT Dharwad)}, 
            city={Dharwad},
            country={India}}
            
\begin{abstract}
Speech produced by individuals with cleft lip and palate (CLP) is often hypernasal (and sometimes breathy) due to structural anomalies, yielding shifts in formant structure that degrade automatic speech recognition (ASR) performance and fairness. Building on evidence that mainstream ASR systems underperform on atypical and disordered speech, we posit that widely used services (e.g., Google Speech-to-Text) exhibit reduced fairness for CLP speech, and we evaluate this claim empirically. To quantify fairness consistently, we introduce a simple fairness score (FS) that trades off overall error and between-group disparity. Despite formant disruptions, mild and moderate CLP speech retains partial spectro-temporal alignment with typical speech, motivating the use of mixing strategies to improve fairness.
We systematically investigated severity-aware \textcolor{black}{mixing} of CLP and normal speech at different severity levels and assessed its effect on fairness. %Three ASR models (GMM–HMM, Whisper, and XLS-R) were evaluated on AIISH (Kannada) and NMCPC (English). The aug improved fairness and reduced WER, from 22.64\% → 18.76\% (GMM–HMM, AIISH) and 28.45\% → 18.89\% (Whisper, NMCPC). 
%GMM–HMM performed best on AIISH, likely reflecting robustness to Kannada child speech, where foundation models struggled. A fairness score quantified relative gains of 17.89\% (AIISH) and 47.50\% (NMCPC).
%We systematically investigate severity-aware aug, wherein CLP speech is combined with normal speech across varying severity levels, and assess its impact on ASR fairness. 
Three ASR models GMM-HMM, Whisper, and XLSR were evaluated on the AIISH (Kannada language) and NMCPC (English language) datasets. A mixing strategy that leverages severity-aware \textcolor{black}{mixing} of CLP and normal speech improves fairness on both English (NMCPC) and Kannada (AIISH) corpora.
%The aug strategy directly improves fairness in the performance of the models on both English (NMCPC) and Kannada (AIISH) speech datasets. 
\textcolor{black}{Notably, the word error rate (WER) decreased from $37.58\%$ to $25.47\%$ (GMM-HMM, AIISH) and from $35.74\%$ to $21.72\%$ (Whisper, NMCPC). The superior performance of GMM-HMM on AIISH is likely attributable to its effectiveness with Kannada children's speech, which poses challenges for foundation models like XLSR and Whisper. FS improved by $32.22\%$ (AIISH) and $39.22\%$ (NMCPC) with the proposed mixing strategy.} These results establish a simple, data-centric mitigation that improves both accuracy and fairness without architectural changes, provide a cross-lingual fairness audit of mainstream ASR on CLP speech, and deliver actionable baselines and metrics that can guide deployment in clinical and accessibility settings.

\end{abstract}

%%Graphical abstract
% \begin{graphicalabstract}
% \includegraphics{grabs}
% \end{graphicalabstract}

%%Research highlights
%\begin{highlights}
%\item This work analyzed the fairness of ASR with CLP speech.

%\item wav2vec2 based ASR techniques have been implemented along with classical techniques to measure ASR and gave an improvement of $00\%$.

%\item Motivated by KALDI based aug for ASR enhancement task, aug is also performed.

%\item The proposed fairness score method was applied to obtain degree of fairness of CLP patients.

% \item Compared to the gold standard unsupervised distance-based approach, the experiments in the Microsoft code-switched (MSCS) dataset show a relatively improved performance of $19.3\%$, $47.3\%$, and $50.7\%$ by using the GMM-UBM, attention, and GAN-based framework, respectively.
%\end{highlights}

\begin{keyword}
%% keywords here, in the form: keyword \sep keyword
Automatic speech recognition \sep XLSR \sep Whisper \sep KALDI \sep Fairness objective score 

%% PACS codes here, in the form: \PACS code \sep code
% \PACS 0000 \sep 1111
% %% MSC codes here, in the form: \MSC code \sep code
% %% or \MSC[2008] code \sep code (2000 is the default)
% \MSC 0000 \sep 1111
\end{keyword}

\end{frontmatter}

%% \linenumbers

%% main text
% \section{Sample Section Title}
% \label{sec:sample1}
\section{\label{sec:1} Introduction}

CLP is a congenital abnormality of the craniofacial region~\cite{evaluationand, methofperceptualassessment, clpnatureandremediation}. Globally, an estimated $192,708$ people were living with CLP in $2019$~\cite{global}. %Globally, the incident of CLP was recorded to be $192,708$ in $2019$~\cite{global}. 
The abnormality in the craniofacial region causes defects in the produced speech~\cite{evaluationand, methofperceptualassessment, clpnatureandremediation,journalofvoice}. Mostly, due to the opening between the lip and nasal cavity, the produced speech is breathy and highly nasalized~\cite{Zajac2011ReliabilityAV}, \cite{Whitehill2004SinglewordII}. Furthermore, the opening between the oral and nasal cavities alters vocal-tract resonances (adding nasal formants and anti-resonances), which manifest as hypernasality and audible nasal air emission and impair the production of high-pressure consonants (stops, fricatives, affricates), thereby reducing speech intelligibility~\cite{evaluationand}, \cite{clpeffectsonspeechandresonance}. Consequently, CLP speech departs from the acoustic–phonetic distribution that mainstream algorithms are trained on: nasal coupling introduces a low-frequency nasal formant and anti-resonances that shift formant frequencies, while breathiness lowers harmonicity and attenuates high-frequency energy in stops/fricatives. These perturbations distort speech, so models trained on normal speech interpret the signal with mismatched priors, degrading interpretive fidelity. %However, to promote equal access to technology, it is necessary to make speech technologies fair for CLP speech.
Ensuring robustness to CLP speech is therefore a prerequisite for safe healthcare communication, equitable educational and civic participation, and responsible, bias-conscious deployment of speech technologies and not simply a matter of equal access.

\begin{figure}[t]
  \centering
  \includegraphics[height=220pt,width=420pt]{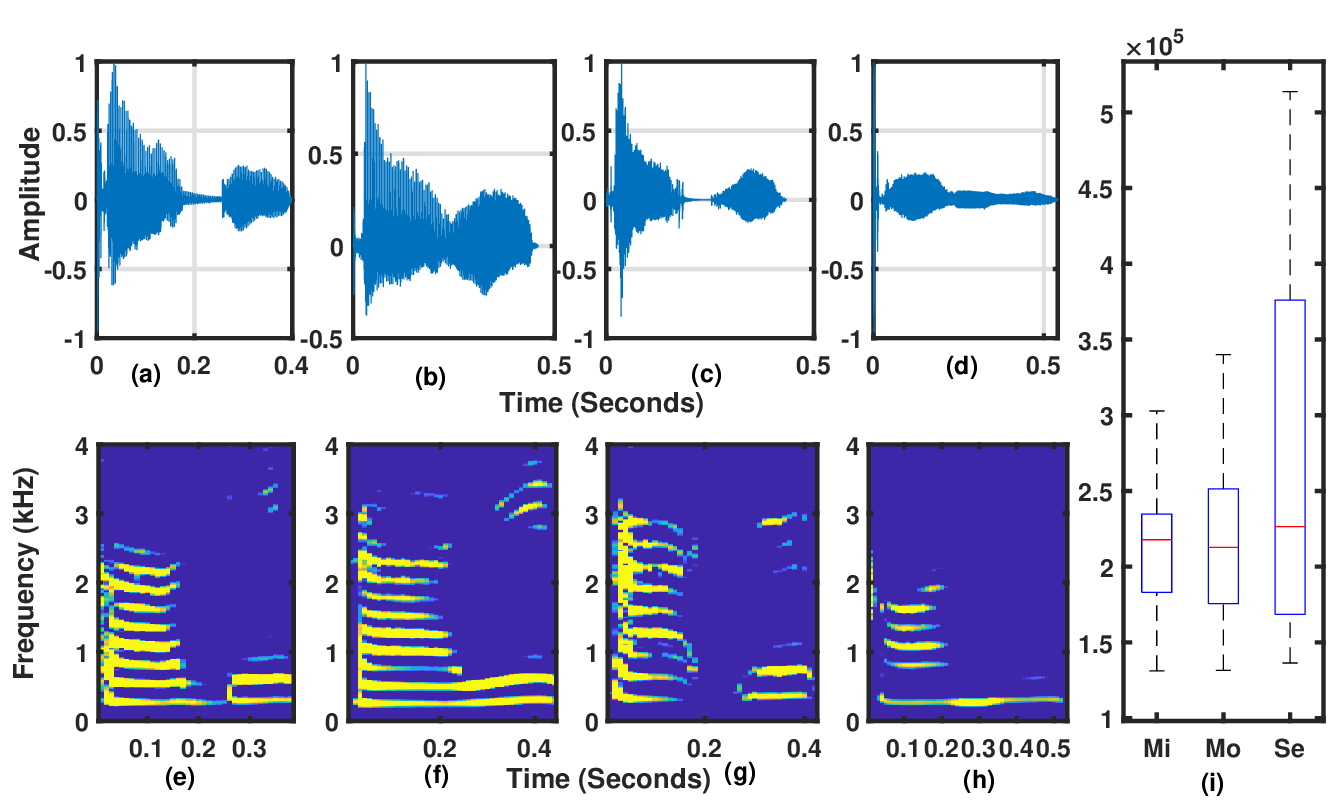}
  \caption{(a--d) Waveforms (\textbf{Amplitude} vs.\ \textbf{Time}) and (e--h) corresponding spectrograms (\textbf{Frequency (kHz)} vs.\ \textbf{Time}) for speech produced by \textbf{normal}, \textbf{mild}, \textbf{moderate}, and \textbf{severe} CLP speakers while producing \textit{kage}. Panel (i) shows the \textbf{DTW distance} ($\times 10^{5}$) \textbf{distribution} between the voiced spectrogram of the normal reference and the mild (Mi), moderate (Mo), and severe (Se) tokens.}
  \label{fig2}
  \vspace{-0.5cm}
\end{figure}

Although prior CLP research has focused on classification and intelligibility enhancement, it often lacks cross-language, child-speech and severity-aware evaluation, relies on surrogate metrics such as phone-decoding accuracy, feature-classifier AUC, \textcolor{black}{PESQ, STOI} rather than end-task ASR WER or listener-rated intelligibility and seldom examines fairness gaps in widely used ASR services. These limitations motivate our study of fairness and mixing CLP speech in Normal across Kannada (AIISH) and English (NMCPC) using GMM-HMM, Whisper and XLSR.
%There exists several attempts in the literature to use CLP speech, mostly to perform classification and intelligibility enhancement. 
The work by Baumann~\textit{et al.} utilizes wav2vec2 embeddings and applies them with machine learning classifiers~\cite{classi_trans} and transformer classifiers perform normal and CLP classification~\cite{ihci}. 
Kalita~\textit{et al.} reported that compensating hypernasality can improve intelligibility and proposed objective measures that combine articulation and hypernasality cues~\cite{intelligibility}. They also introduced a glottal-activity based hypernasality score~\cite{glottis} and employed deep models with Gaussian posteriorgrams to estimate hypernasality~\cite{gauss}.
%It is also reported in~\cite{intelligibility} that the intelligibility of the speech can be improved by compensating the hypernasality. The work in~\cite{glottis}, proposed a technique using the glottal activity region to compute the hypernasality score, while in~\cite{gauss} deep learning approaches are used to estimate hypernasality. In~\cite{trills}, the work proposed a procedure to detect the stop consonant and trill sounds and then replace them to enhance the speech perception. 
At the segment level, Vikram~\textit{et al.} analyzed misarticulated trills in CLP speech~\cite{trills}. In~\cite{sensitivity}  a MaskCycleGAN-based enhancement was demonstrated that improved phone decoding with a GMM–HMM backend. Yao~\textit{et al.}~\cite{applied} proposed an IP-encoded Whale Optimization Algorithm to automatically evolve variable-length DCNNs for pathological-speech classification including CLP and thereby improving CLP recognition via optimized architectures. Fu et al.~\cite{cls_clp} introduced the Vocal Tract Area Spectrum (VTAS) and a set of VTAS-based acoustic features capturing \textcolor{black}{articulatory variation and complexity} and used an AdaBoost classifier to automatically detect pharyngeal fricatives in CLP speech.
%In~\cite{trills}, stop and trill segments are identified and substituted to improve perception. In~\cite{sensitivity}, a MaskCycleGAN-driven enhancement of CLP speech yields better phone decoding performance with a Gaussian mixture model-hidden Markov model (GMM-HMM) backend.
Further, several works also attempt to enhance the intelligibility of CLP speech using the generative end-to-end deep learning frameworks~\cite{DL1,DL2}.  However, as per our knowledge, no work exists in the literature that studies the fairness of ASR in CLP speech. 
Figure~\ref{fig2} illustrates the speech signal, spectrogram, and formant contours for speech produced by normal, mild, moderate, and severely affected CLP subjects. The figure indicates a progressive shift in formant locations from normal to severe, likely due to the increasing degree of oral-nasal tract opening. However, while the shape of the formant contours remains largely intact in normal and mild speech, it gradually deteriorates in moderate and severe cases. This suggests that ASR performance may decline due to formant frequency shifts.

\begin{comment}

\begin{figure}
	\centering
	\includegraphics[height= 220pt,width=420pt]{Interspeech_mild_mod_sev.eps}
	\caption{ (a-h) Speech signal produced by Normal, mild, moderate, and severe subjects, and their corresponding spectrogram, and (i) showing the DTW distance distribution between the voiced spectrogram of normal to mild (Mi), moderate (Mo), and severe (Se), respectively, while producing the sound \textbf{\emph{"kage"}}. }
	\label{fig2}
 \vspace{-0.5 cm}
\end{figure}

\end{comment}

% \begin{figure}
% 	\centering
% 	\includegraphics[height= 180pt,width=300pt]{interspeech24_clp.jpg}
% 	\caption{ Speech signal produced by Normal, mild, moderate and severe subjects, and their corresponding spectrogram and formant contour, while producing the sound \textbf{\emph{"kage"}}.}
% 	\label{fig1}
%  \vspace{-0.5 cm}
% \end{figure}

This study investigates the performance gap in ASR between typical (normal) speech and speech affected by CLP, using the state-of-the-art Google ASR API as a baseline. To further explore the similarities and differences between normal and CLP speech, we perform cross-training and cross-testing experiments. These experiments utilize three established ASR approaches: (1) Gaussian Mixture Model - Hidden Markov Model (GMM-HMM), (2) cross-lingual unsupervised speech representation (XLSR), and (3) Whisper, a large-scale weakly supervised sequence-to-sequence ASR model developed by OpenAI. We conduct experiments on two datasets: (1) AIISH, consisting primarily of Kannada children's speech, and (2) NMCPC, which comprises English children's speech. The selected ASR approaches reflect consideration of the available data volume. GMM-HMM is a traditional method known for its effectiveness in low-resource settings. XLSR-wav2vec2 represents a recently popular unsupervised approach, while Whisper is based on weakly supervised large-scale pretraining. Despite limited dataset availability for CLP speech, AIISH and NMCPC provide a representative comparison across languages and recording conditions. While the primary focus is on fairness in ASR performance, this study also aims to assess which ASR approach performs optimally on each dataset, considering their linguistic differences. Notably, foundation models like XLSR and Whisper have been predominantly trained on English, with significantly less exposure to Kannada.

Research on CLP speech has primarily focused on hypernasality detection, intelligibility assessment and speech enhancement. Hypernasality, a key characteristic of CLP speech, has been extensively studied using various signal processing techniques. Early works such as Dubey~\textit{et al.} introduced zero-time windowing (ZTW) and homomorphic cepstral front ends for hypernasality detection, pairing these features with SVMs for CLP/normal classification \cite{Dubey2016ZeroTW,Dubey2018HypernasalityDU}. Nikitha ~\textit{et al.} analyzed formant-based vowel-space reductions in children with CLP \cite{hypernasalityseveritykalita}. VikramC.~\textit{et al.} proposed a posterior-probability framework to estimate hypernasality scores, addressing limitations of perceptual ratings \cite{estimationhypernasality}. Mathad~\textit{et al.} developed an attention-based BLSTM that improved automatic hypernasality prediction \cite{attentionhypernasality}. Along the feature-engineering line, Dubey~\textit{et al.} designed pitch-adaptive front ends and constant-Q cepstral coefficients (CQCC) to boost detection accuracy \cite{Dubey2018PitchAdaptiveFF,Dubey2019HypernasalitySD}, and later explored sinusoidal-model features and combined vocal-tract/residual cues for more robust detection \cite{Dubey2020SinusoidalMH,Dubey2019HypernasalitySD}. More recent studies improved classification using advanced spectral/articulatory features \cite{DubeyDetection} and latent representations from pre-trained Wav2Vec2 \cite{improving}.

Intelligibility assessment studies have employed a variety of methodologies to estimate speech intelligibility, including Gaussian posteriorgrams \cite{gauss}, dynamic time warping (DTW) \cite{selfsimilaritysishir} and regression models \cite{objectiveassessmentsishir}. These approaches have demonstrated strong correlations with human perceptual ratings, underscoring their effectiveness for objective intelligibility assessment. For instance, studies like \cite{sprotima} have analyzed spectral moments of fricatives in CLP speech, while \cite{nazalizedstops} leveraged epoch-synchronous features to detect nasalized voiced stops. In the realm of speech enhancement, significant efforts have been made to improve intelligibility using both signal processing and deep learning techniques. Early works, such as \cite{stransientprotima}, focused on modifying phoneme transitions, whereas \cite{tempspecprotima} proposed enhancements for hypernasal speech by altering vocal tract system characteristics. Misarticulated fricative and stop consonants were addressed in \cite{fricativeprotima} and \cite{eventbasedmisarticulatedstops} through spectral transformations. More recent advancements have utilized deep learning-based methods, such as CycleGAN in \cite{cyclegan1}, which significantly improved word error rate (WER) in ASR evaluations for CLP speech. Additionally, \cite{dataaugpro} demonstrated enhanced CLP speech recognition performance using data \textcolor{black}{mixing} techniques like vocal tract length perturbation and reverberation. A summary of existing works using CLP speech is presented in Table~\ref{summary}.

Although prior studies have primarily focused on intelligibility assessment and enhancement, there is a lack of research evaluating the performance and fairness of recent ASR methods on CLP speech. To address this gap, we systematically analyze ASR performance on CLP speech and propose a severity-aware data \textcolor{black}{mixing} strategy to improve the fairness of ASR systems.

Moreover, Figure~\ref{fig2} shows that the dynamics of the resonances (highlighted as yellow bands) in the spectrogram become increasingly distorted as the severity level rises. Figures~\ref{fig2}(e) and (f), in particular, retain nearly intact spectral resonance patterns. These observations motivate us to investigate severity-aware data \textcolor{black}{mixing} for evaluating the performance and fairness of ASR systems. In this work, we perform severity speech \textcolor{black}{mixing} in a controlled manner and analyze the performance of ASR approaches under both typical and CLP speech conditions, with a specific focus on fairness.

The main contributions of this work are summarized as follows:   \begin{enumerate}
   \item We evaluate the fairness of publicly available ASR systems on CLP speech and demonstrate their degraded performance under such conditions.

\item We propose a severity-aware data \textcolor{black}{mixing} method to improve ASR fairness across different levels of CLP speech severity \textcolor{black}{by mixing with real CLP data severities in equal proportion with normal speech}.

\item We analyze effective ASR approaches for CLP speech in challenging scenarios, including children's speech and low-resource language settings.

%While previous studies have considered the levels of severity in CLP speech, our work takes a stepwise approach by progressively merging severity levels to analyze their collective impact.

%\item A fairness score is introduced to quantify and assess improvements in ASR performance for CLP speech.
\end{enumerate}

\begin{table}[]
\centering
\caption{Summary of work done so far in CLP }
\label{summary}
\resizebox{\textwidth}{!}{%
\begin{tabular}{|l|l|l|l|l|}
\hline
\multicolumn{1}{|c|}{\textbf{Sl.}} &
  \multicolumn{1}{c|}{\textbf{Ref, Year}} &
  \multicolumn{1}{c|}{\textbf{Task}} &
  \multicolumn{1}{c|}{\textbf{Dataset}} &
  \multicolumn{1}{c|}{\textbf{Method}} \\ \hline
1 &
  \cite{Dubey2016ZeroTW} , 2016 &
  \begin{tabular}{p{3cm}}Hypernasality Detection\end{tabular} &
  \begin{tabular}{p{1.7cm}} AIISH \end{tabular} &
  \begin{tabular}{p{17cm}}Based on presence/absence of extra nasal peak in low, high and voiced consonants in the HNGD spectrum, the severity rating of the hypernasal speech can be decided for /a/ and /i/ vowels and voice consonants /b/, /d/ or /g/  \end{tabular} \\ \hline

2 &
  \cite{stransientprotima} , 2018 &
  \begin{tabular}{p{3cm}} Intelligibility improvement\end{tabular}  &
  \begin{tabular}{p{1.7cm}} AIISH \end{tabular} &
  \begin{tabular}{p{17cm}}This study enhances intelligibility of /s/ substituted by a glottal stop by inserting sustained portions and modifying transitions using 2D-DCT projections onto normal speech SVD vectors.
\end{tabular} \\ \hline

3 &
  \cite{trills} , 2018 &
  \begin{tabular}{p{3cm}} Misarticulated trills analysis\end{tabular}  &
  \begin{tabular}{p{1.7cm}} AIISH \end{tabular} &
  \begin{tabular}{p{17cm}}Acoustic analysis of misarticulated trills in CLP children using glottal and vocal tract features shows significant differences from normal speech. A DTW-based system using trill-specific features outperforms MFCCs in detecting misarticulations. \end{tabular} \\ \hline
  
4 &
  \cite{glottalactivityvikram} , 2018 &
  \begin{tabular}{p{3cm}} Glottal Activity Errors Detection  in Stops\end{tabular} &
  \begin{tabular}{p{1.7cm}} AIISH \end{tabular} &
  \begin{tabular}{p{17cm}}The proposed algorithm detects glottal activity errors (GAE) in stop consonant production of CLP speakers using low-frequency voiced consonant evidence from zero-frequency filtering (ZFFS) and band-pass filtering (BPFS) %\\ This can assist speech-language pathologists (SLPs) in assessing articulation errors in CLP speech.
  \end{tabular} \\ \hline

5 &
  \cite{estimationhypernasality} , 2018 &
  \begin{tabular}{p{3cm}} Hypernasality estimation\end{tabular}  &
  \begin{tabular}{p{1.7cm}} AIISH \end{tabular} &
  \begin{tabular}{p{17cm}} Motivated by the functionality of nasometer, a posterior probability-based approach is proposed here which estimates hypernasality scores using MFCCs from glottal regions. DNN outperforms GMM and also nasometer. \end{tabular} \\ \hline

6 &
  \cite{glottis} , 2018 &
  \begin{tabular}{p{3cm}}Intelligibility assessment\end{tabular} &
  \begin{tabular}{p{1.7cm}} AIISH \end{tabular} &
  \begin{tabular}{p{17cm}} This study analyzes CLP speech intelligibility using glottal landmarks (g LMs) and acoustic features, showing that Mel-2DDCT-based GMMs outperform MFCCs by better capturing abrupt transitions and correlating with perceptual ratings.\end{tabular} \\ \hline
7 &
  \cite{gauss} , 2018 &
  \begin{tabular}{p{3cm}} Intelligibility assessment\end{tabular} &
  \begin{tabular}{p{1.7cm}} AIISH \end{tabular} &
  \begin{tabular}{p{17cm}}This study uses GP-based speech representation and DTW distance to assess CLP child speech intelligibility, showing that pitch-normalized Mel-2D-DCT features best correlate with SLP perceptual ratings, outperforming MFCCs and LP-2D-DCT features.\end{tabular} \\ \hline
8 &
  \cite{selfsimilaritysishir} , 2018 &
  \begin{tabular}{p{3cm}} Intelligibility Assessment\end{tabular} &
  \begin{tabular}{p{1.7cm}} AIISH \end{tabular} &
  \begin{tabular}{p{17cm}}This study proposes an SSM-based unsupervised framework for estimating CLP children's speech intelligibility, showing that GP-based SSMs outperform MFCCs and DTW in correlating with perceptual ratings and discriminating intelligibility groups. \end{tabular} \\ \hline
9 &
  \cite{Dubey2018HypernasalityDU} , 2018 &
  \begin{tabular}{p{3cm}}Hypernasality Detection \& assessment\end{tabular} &
  \begin{tabular}{p{1.7cm}} AIISH \end{tabular} &
  \begin{tabular}{p{17cm}} This study introduces the HNGDF cepstral feature for hypernasality detection, showing superior accuracy over EDM features, with further improvement when combined with MFCCs, making it promising for hypernasality severity analysis.\end{tabular} \\ \hline
10 &
  \cite{Dubey2018PitchAdaptiveFF} , 2018 &
  \begin{tabular}{p{3cm}} Hypernasality Detection\end{tabular} &
  \begin{tabular}{p{1.7cm}} AIISH \end{tabular} &
  \begin{tabular}{p{17cm}}This study proposes Pitch-Adaptive MFCC (PAMFCC) for hypernasality detection, improving low-frequency nasality cue capture and achieving higher classification accuracy than MFCCs by mitigating pitch harmonics effects in CLP speech.\end{tabular} \\ \hline
  
11 &
  \cite{sprotima} , 2019 &
  \begin{tabular}{p{3cm}}Study of voiceless sibliant fricatives\end{tabular} &
  \begin{tabular}{p{1.7cm}} AIISH \end{tabular} &
  \begin{tabular}{p{17cm}} This work analyzes NAE-affected voiceless sibilant fricatives in Kannada, showing spectral deviations due to VPD. An SVM classifier using spectral moments and peak ERBN-number achieves high accuracy in detecting NAE-distorted fricatives.\end{tabular} \\ \hline

12 &
  \cite{nazalizedstops} , 2019 &
  \begin{tabular}{p{3cm}} Segmentation \& detection of nasalized voiced stops\end{tabular} &
  \begin{tabular}{p{1.7cm}} AIISH \end{tabular} &
  \begin{tabular}{p{17cm}}This study proposes an automatic segmentation and detection algorithm for nasalized voiced stops in CP speech, using glottal activity and spectral features to enhance SVM-based classification, outperforming HMM-based segmentation and MFCCs.\end{tabular} \\ \hline
13 &
  \cite{objectiveassessmentsishir} , 2019 &
  \begin{tabular}{p{3cm}} Composite measure of speech intelligibility\end{tabular} &
  \begin{tabular}{p{1.7cm}} AIISH \end{tabular} &
  \begin{tabular}{p{17cm}} This study proposes a composite intelligibility measure for CLP speech using articulation and hypernasality features, with SVR achieving the best prediction of PCC scores using wM2DDCT, wMFCC, and gMFCC features. \end{tabular} \\ \hline
14 &
  \cite{DubeyDetection} , 2019 &
  \begin{tabular}{p{3cm}} Hypernasality Detection\end{tabular}  &
  \begin{tabular}{p{1.7cm}} AIISH \end{tabular} &
  \begin{tabular}{p{17cm}} This study detects hypernasality using VTC, PSR, and SMAC features, capturing spectral distortions in vowels. SVM classification with combined features outperforms baselines for both detection and severity classification.\end{tabular} \\ \hline

15 & \cite{Dubey2019HypernasalitySD} , 2019 &
  \begin{tabular}{p{3cm}}Hypernasality severity Detection\end{tabular} &
  \begin{tabular}{p{1.7cm}} AIISH \end{tabular} &
  \begin{tabular}{p{17cm}}This study detects hypernasality severity in /i/ and /u/ vowels using CQCC features, which capture nasal formant variations more effectively than MFCCs, improving accuracy but with challenges in mild case classification.  \end{tabular} \\ \hline
16 &
  \cite{Dubey2020SinusoidalMH} , 2020 &
  \begin{tabular}{p{3cm}} Hypernasality Detection \end{tabular} &
  \begin{tabular}{p{1.7cm}} AIISH \end{tabular} &
  \begin{tabular}{p{17cm}}This study proposes NHA, HAR, and PHF features for hypernasality detection using a sinusoidal speech model, with SVM classification showing that their combination outperforms individual features and baseline methods. \end{tabular} \\ \hline
17 &
  \cite{tempspecprotima} , 2020 &
  \begin{tabular}{p{3cm}} Enhancement of Vowels \end{tabular} &
  \begin{tabular}{p{1.7cm}} AIISH \end{tabular} &
  \begin{tabular}{p{17cm}}This study explores hypernasal speech enhancement using XLP residual modification, vocal tract system modification, and their combination, with evaluations showing the combined approach most effectively reduces nasalization.%, \\ outperforming individual modifications.. 
  \end{tabular} \\ \hline

18 &
  \cite{vop} , 2020 &
  \begin{tabular}{p{3cm}} Detection of misarticulated stops\end{tabular} &
  \begin{tabular}{p{1.7cm}} AIISH \end{tabular} &
  \begin{tabular}{p{17cm}}This study segments CV transitions in CLP speech using VOPs and SPF-based 2D-DCT features, with SVM classification outperforming STFT-based 2D-DCT, MFCCs, and HMM models in detecting misarticulated stops. \end{tabular} \\ \hline

19 &
  \cite{fricativeprotima} , 2021 &
  \begin{tabular}{p{3cm}}Modification of fricatives \end{tabular} &
  \begin{tabular}{p{1.7cm}} AIISH \end{tabular} &
  \begin{tabular}{p{17cm}} This study modifies misarticulated /s/ in CLP speech using spectral adjustments and synthesized insertions, improving spectral similarity and intelligibility, though MOS scores remain below normal. \end{tabular} 
  \\ \hline

20 &
  \cite{eventbasedmisarticulatedstops} , 2021 &
  \begin{tabular}{p{3cm}} Misarticulated stops Enhancement\end{tabular} &
  \begin{tabular}{p{1.7cm}} AIISH \end{tabular} &
  \begin{tabular}{p{17cm}}This study enhances CLP speech intelligibility by modifying misarticulated stops using NMF-based spectral transformation, improving detection and perceptual similarity, though MOS ratings suggest room for further quality enhancement. \end{tabular} \\ \hline
21&
  \cite{objassesment} , 2021 &
  \begin{tabular}{p{3cm}} Hypernasality Assessment\end{tabular}  &
    \begin{tabular}{p{1.7cm}}Americleft,\\ NMCPC\end{tabular}  &
  \begin{tabular}{p{17cm}} This study proposes OHM, a DNN-based hypernasality assessment metric trained on healthy speech, achieving high correlation with expert ratings and sensitivity to mild hypernasality, performing comparably to clinicians.\end{tabular} \\ \hline
22 &
  \cite{cyclegan1} , 2021 &
  \begin{tabular}{p{3cm}} Intelligibility Enhancement\end{tabular} &
  \begin{tabular}{p{1.7cm}} AIISH \end{tabular} & 
  \begin{tabular}{p{17cm}} This study uses CycleGAN to enhance CLP children's speech intelligibility, with ASR and subjective evaluations confirming improvements, benefiting speech-controlled device usability and therapy outcomes.\end{tabular} \\ \hline
23 &
  \cite{dataaugpro} , 2021 &
  \begin{tabular}{p{3cm}} Data mixing for improving CLP ASR\end{tabular} &
  \begin{tabular}{p{1.7cm}} AIISH \end{tabular} & 
   \begin{tabular}{p{17cm}}This study explores data mixing for CLP speech recognition, with CycleGAN, VTLP, and reverberation showing the best improvements, significantly reducing phone error rates. \end{tabular} \\ \hline

24 &
  \cite{freq} , 2021 &
  \begin{tabular}{p{3cm}} Data mixing based on Frequency Warping\end{tabular} & 
   \begin{tabular}{p{1.7cm}} ATR Japanese speech \end{tabular} & 
   \begin{tabular}{p{17cm}}This paper proposes frequency warping for data mixing in CLP speech ASR, enhancing robustness to formant fluctuations and improving accuracy when combined with unsupervised learning, outperforming SpecAugment.\end{tabular} \\ \hline

25 &
  \cite{sensitivity} , 2022 &
  \begin{tabular}{p{3cm}} Intelligibility Enhancement \end{tabular} & 
   \begin{tabular}{p{1.7cm}} AIISH \end{tabular} & 
   \begin{tabular}{p{17cm}}This paper proposes MaskCycleGAN for data mixing in CLP speech ASR and obtained better WER outperforming CycleGAN.\end{tabular} \\ \hline

26 &
  \cite{improving} , 2022 &
  \begin{tabular}{p{3cm}} Hypernasality Estimation \end{tabular} &
  \begin{tabular}{p{1.7cm}}CNH-CLP,\\ NMCPC\end{tabular} &
   %CNH-CLP, NMCPC & 
   \begin{tabular}{p{17cm}} 
   This study improves hypernasality estimation by fine-tuning a pre-trained ASR encoder, leveraging larger ASR datasets and text labels for better feature extraction, achieving superior performance on cleft palate datasets.
   \end{tabular} \\ \hline

27 &
  \cite{influence} , 2023 &
  \begin{tabular}{p{3cm}}Classification of CLP \& Normal\end{tabular} & \begin{tabular}{p{1.7cm}}
   Erlangen-CLP \end{tabular}& 
   \begin{tabular}{p{17cm}} 
  Classification between
CLP and healthy voices with latent representations from the lower and middle \\encoder layers of a pre-trained wav2vec 2.0 system, reached an accuracy of 100\%. 
   \end{tabular} \\ \hline

27 &
  \cite{ihci} , 2023 &
  \begin{tabular}{p{3cm}} Classification of CLP \& Normal using transformers \end{tabular} & \begin{tabular}{p{1.7cm}}
   AIISH,\\ NMCPC \end{tabular}& 
   \begin{tabular}{p{17cm}} 
 This study fine-tunes pretrained transformer models on CLP speech, showing superior classification performance, with DistilHuBERT achieving near 100\% accuracy. 
   \end{tabular} \\ \hline

\textbf{29} &
  \textbf{Ours, 2025} &
  \textbf{\begin{tabular}{p{3cm}}Fairness of ASR  \\ in CLP Speech \end{tabular}} &
\begin{tabular}{p{1.7cm}}\textbf{ AIISH},\\ \textbf{NMCPC}\end{tabular} &
% \textbf{ AIISH, NMCPC} & 
   \begin{tabular}{p{17cm}} 
   This study examines ASR fairness for CLP speech, evaluating GMM-HMM, Whisper, and XLSR systems and exploring the impact of \textcolor{black}{mixing} CLP speech with varying severity levels of normal speech.
   \end{tabular} \\ \hline
\end{tabular}

}

\end{table}

%A summary of these works is provided in Table~\ref{summary}. However, despite these advancements, none of the existing studies address the fairness of ASR systems for CLP speech. In this work, we aim to bridge this gap by evaluating the fairness of ASR systems for CLP speech and exploring the application of data aug tailored to the severity levels of CLP speech. 
%This approach seeks to ensure equitable performance of ASR systems across varying degrees of speech impairment, thereby promoting inclusivity in speech technology.

\section{Database setup}

%This section outlines the characteristics of the speech datasets used to evaluate ASR fairness across normal and CLP speech conditions.%This section provides a brief description of the database used in this study. 
Given the scarcity of publicly available CLP speech corpora on CLP speech~\cite{jawed, erlangen, ultrasuite}, we used two datasets to perform the experiments. Initially, the analysis is performed with All India Institute of Speech and Hearing (AIISH)~\cite{hypernasality}, and then, for generalization purposes, some experiments are repeated with the New Mexico Cleft Palate Centre (NMCPC) dataset~\cite{jawed}. 
%We selected  datasets because high-fidelity, clinician-annotated CLP corpora are scarce, and both datasets provide expert SLP labels (CLP diagnosis/severity) together with parallel normal-control recordings collected under comparable protocols. This pairing allowed like-for-like comparisons of normal vs. CLP speech and supported our severity-aware aug design. Using two languages also enabled a cross-lingual generalization check. We conducted the primary analyses on AIISH and repeated key experiments on NMCPC to assess robustness.
We selected AIISH and NMCPC datasets because they include expert annotations from speech-language pathologists, with samples categorized by severity. Hence these existing datasets were utilized and speaker-dependent partitioning is employed for our experiments.
The former is in the Kannada language, whereas the latter is in English.  Both datasets are randomly divided into training and testing partitions with $80:20$ ratio. After that, the training partition is further partitioned to $80:20$ to form the training and development set. 

\textbf{AIISH dataset:} The dataset is collected from the All India Institute of Speech and Hearing, India.  It consists of $60$ speakers with $31$ normal and $29$ CLP speakers. Out of them, $19$  and $12$ are normal female and male speakers respectively, and $9$ and $20$ are CLP female and male speakers, respectively. The dataset is collected with $19$ unique utterances.  Each sentence has a maximum of $3$ words. The participants are native Kannada speakers, who are within the age group $7-12$ years. They did not have any other congenital syndromes like hearing impairment. The detailed statistics of the dataset are given in Table~\ref{dataset_description}. %Out of a total of $2,726$ utterances, the dataset is partitioned as follows: training ($1,731$ utterances – $1,106$ normal, $625$ CLP), development ($429$ utterances – $278$ normal, $151$ CLP), and evaluation ($566$ utterances – $357$ normal, $209$ CLP).

%\vspace{0.1 cm}\\
%The speech signals are processed in frames of $20$ ms with a shift of $10$ ms and characterized by $1024$ fast-Fourier transform (FFT) points.

\begin{table}[]
\centering
\caption{Description of AIISH and NMCPC~\cite{jawed} datasets}
\begin{tabular}{|c|c|c|c|c|c|}
\hline
\begin{tabular}[c]{@{}c@{}}Dataset\end{tabular} &
\begin{tabular}[c]{@{}c@{}}Type of\\ audio Data\end{tabular} &
\begin{tabular}[c]{@{}c@{}}Severity of\\ Speaker\end{tabular} & Subjects & Utterances & \textbf{Total (Normal / CLP)} \\ \hline

\multirow{4}{*}{AIISH} & Normal  & -        & 31 & 1741 & \multirow{4}{*}{1741 / 985} \\ \cline{2-5} 
                       & \multirow{3}{*}{CLP} & Mild     & 14 & 473 &  \\ \cline{3-5} 
                       &                      & Moderate & 11 & 379 &  \\ \cline{3-5}  
                       &                      & Severe   & 4  & 133 &  \\ \hline

\multirow{4}{*}{NMCPC} & Normal  & -        & 24 & 439 & \multirow{4}{*}{439 / 1024} \\ \cline{2-5} 
                       & \multirow{3}{*}{CLP} & Mild     & 11 & 385 &  \\ \cline{3-5} 
                       &                      & Moderate & 14 & 324 &  \\ \cline{3-5}  
                       &                      & Severe   & 16 & 315 &  \\ \hline

\end{tabular}
\label{dataset_description}
\end{table}

\textbf{NMCPC dataset:} The dataset is collected at the New Mexico Cleft Palate Centre. It has a total of 65 speakers and consists of speech utterances from $41$ CLP speakers ($22$ male and $19$ females) and $24$ normal speakers ($20$ male and $4$ females). The dataset consists of $76$ unique utterances.  Each sentence has a maximum of $5$ words. The age group of the speakers is $9-13$ years. The CLP speakers were classified into mild, moderate and severe.  The detailed statistics of the dataset are given in Table~\ref{dataset_description} \cite{jawed}.

%Out of a total of $1,463$ utterances, the dataset is divided into training ($929$ utterances – $649$ normal, $280$ CLP), development ($235$ utterances – $165$ normal, $70$ CLP), and evaluation ($299$ utterances – $89$ normal, $210$ CLP).

\textcolor{black}{Both the AIISH and NMCPC data sets had their training and evaluation set  formed such that the speakers therein are disjoint from those in the test set. Therefore the external test set only contains unseen speakers.
To ensure that performance differences were attributable to the composition of the training data rather than to variations in the total number of training utterances, a controlled sampling strategy was adopted for both the NMCPC and AIISH datasets. For each dataset, the total size of every training configuration was fixed to the number of normal-speech training files available in that dataset. The samples were distributed as evenly as possible among the severity groups included in each configuration. For the NMCPC dataset, the available training data comprised 280 normal, 246 mild, 204 moderate and 199 severe speech files. Since the normal-speech partition contained 280 files, the total number of training files was fixed at 280 for all controlled experiments. The Normal-only configuration therefore contained 280 normal-speech files. The Normal+Mild configuration contained 140 normal and 140 mild files. The Normal+Mild+Moderate configuration contained 93 normal, 93 mild, and 94 moderate files, while the Normal+Mild+Moderate+Severe configuration contained 70 files from each severity group. The CLP-only reference configuration was formed using 93 mild, 93 moderate, and 94 severe files, resulting in the same total of 280 training files. In addition, 70 development files were merged from the corresponding partitions. The evaluation set contained 264 files, with 66 files each from the normal, mild, moderate, and severe groups.
For the AIISH dataset, the available training data comprised 304 normal, 302 mild, 247 moderate, and 76 severe speech files. Accordingly, the total number of training files was fixed at 304 for all controlled configurations. The Normal-only configuration contained 304 normal-speech files. The Normal+Mild configuration contained 152 normal and 152 mild files. The Normal+Mild+Moderate configuration contained 101 normal, 101 mild, and 102 moderate files. For the Normal+Mild+Moderate+Severe configuration, 76 files were selected from each group, resulting in a total of 304 files. The CLP-only reference configuration contained 101 mild, 101 moderate, and 102 severe files, again yielding 304 files in total. 76 development files were merged from the relevant partitions. The evaluation set consisted of 152 files, with 38 files from each of the four severity groups. As in the NMCPC setup, 70 development files were merged from the four partitions. The evaluation set consisted of 264 files, with 66 files from each of the four severity groups.
This controlled design maintained an equal training-set size across all data-composition conditions while varying only the proportion and diversity of the severity groups. It therefore enabled a fair comparison of the effect of severity-aware data inclusion on recognition accuracy and fairness.}

%The dataset is divided into two main parts: 80\% is allocated to the Training Set and the remaining 20\% of the dataset is set aside as the Evaluation Set, which is used to assess the model's performance on unseen data after training. The training set is further split into two subsets: 64\% is used as the Training Subset for training the model, and 16\% is designated as the Dev Subset, which is used for model validation during training. 
%In AIISH, out of total 985 CLP utterances, 625 are used for training, 151 for validation and 209 for evaluation. For Normal utterances, out of 1741 audio files, 1106 are used for training, 278 for validation and 357 for evaluation. In NMCPC, out of total 1024 CLP utterances, 649 are used for training, 165 for validation and 210 for evaluation. For Normal utterances, out of 439 audio files, 280 are used for training, 68 for validation and 89 for evaluation.

\section{Fairness of publicly available Google application programming interface (API) ASR in CLP speech}
\label{ob_study}

This section discusses the performance of publicly available Google API ASR and its fairness when used with CLP speech.

%is to discuss the ASR performances that are available for public use.

\subsection{Google API ASR performance in CLP speech}
We used Google’s~\footnote{ \url{https://cloud.google.com/speech-to-text}} publicly available ASR (English for NMCPC, Kannada for AIISH) to evaluate both datasets on their held-out evaluation splits. WER was computed by comparing the ASR hypotheses with the human reference transcriptions. 

On AIISH, WER was $93\%$ for the normal set and $98.94\%$ for the CLP set; on NMCPC, WER was $30.21\%$ (normal) and $74.27\%$ (CLP). The WER values are summarized in Table~\ref{fairness_nmcpc} ~\cite{jawed}.
%We used Google’s~\footnote{ \url{https://cloud.google.com/speech-to-text}} publicly available ASR (English for NMCPC, Kannada for AIISH) to score both datasets. Results are reported on the held-out evaluation splits, and the WER values are summarized in Table~\ref{fairness_nmcpc} ~\cite{jawed}.
%We use publicly available English and Kannada ASR from Google~\footnote{ \url{https://cloud.google.com/speech-to-text}} to evaluate the performance of the NMCPC and AIISH datasets, respectively. We assess performance on the held-out evaluation splits of both datasets. The performance obtained in terms of WER is tabulated in Table~\ref{fairness_nmcpc} ~\cite{jawed}.
%We use the evaluation set of both datasets to evaluate performance. 
%After decoding, the WER performance was evaluated by comparing them with the ground truth text transcription. 
%The performance obtained in terms of WER for AIISH dataset is $93\%$ and $98.94\%$ in the normal and CLP test sets, respectively, and for NMCPC dataset $30.21\%$ and $74.27\%$ in the normal and CLP test sets, respectively, for the same test set utterances. 
The performance degradation from $93\%$ to $98.94\%$ for AIISH and $30.21\%$ to $74.27\%$ for NMCPC justifies the claim that the fairness of the ASR system is compromised in CLP speech. %However, it is worth noting that the obtained WER of $30.21\%$ in the normal set could be due to the test utterances being from children. 
These findings emphasize the significant performance gap in ASR systems when processing CLP speech, highlighting concerns about fairness and accessibility. The substantial increase in WER for CLP utterances suggests that current ASR models struggle with disordered speech, necessitating targeted improvements in acoustic modeling and adaptation techniques. Furthermore, it should be noted that the relatively high WER for normal speech specifically for AIISH can be attributed to the challenges of child speech recognition~\cite{children}.

%This highlights the need for child-inclusive ASR training.

%\begin{table}[ht]
%\centering
%\captionsetup{width=\textwidth}
%\caption{Fairness of ASR available publicly through Google API, $W_C$, $W_N$, and $\gamma$ are the obtained WER from CLP and normal test utterances and fairness score, respectively for AIISH datset.}
%\label{fairness_aiish}
%\begin{tabular}{|c|c|c|}
%\hline
%\textbf{$W_N \downarrow$} & \textbf{$W_C \downarrow$} & \textbf{$\gamma \downarrow$} \\ \hline
%107.71    & 139.75 &  -114.56   \\ \hline
%\end{tabular}
%\end{table}

\begin{table}[ht]
\centering
\caption{Fairness of ASR available publicly through Google API, $W_C$, $W_N$ are the obtained WER from CLP and normal test utterances, respectively for AIISH \& NMCPC datsets.}
\label{fairness_nmcpc}
\begin{tabular}{|c|c|c|}
\hline
Dataset & \textbf{$W_N \downarrow$} & \textbf{$W_C \downarrow$} \\ \hline
AIISH & 93.00    & 98.94    \\ \hline
NMCPC & 30.21    & 74.27    \\ 
\hline
\end{tabular}
\end{table}

%\begin{table}[ht]
%\centering
%\captionsetup{width=\textwidth}
%\caption{Fairness of ASR available publicly through Google API, $W_C$, $W_N$, and $\gamma$ are the obtained WER from CLP and normal test utterances and fairness score, respectively for NMCPC datset.}
%\label{fairness_nmcpc}
%\begin{tabular}{|c|c|c|}
%\hline
%\textbf{$W_N \downarrow$} & \textbf{$W_C \downarrow$} & \textbf{$\gamma \downarrow$} \\ \hline
%30.21    & 74.27 &  -48.15   \\ %\hline
%\end{tabular}
%\end{table}

%The ASR system developed in this work is as per the best of our knowledge our work is the first where transformer based technique is applied for enhancement purpose and with promising results which can be seen in the following sections.

\subsection{Fairness as a metric}\label{ob_study2}

Inspired by the work of Howard et al.~\cite{fairness1} and Liang et al.~\cite{fairness2}, we use fairness metrics to observe the degree of fairness of the ASR system. The Fairness Score (FS) is defined as a weighted combination of the negative average error rate and the error disparity between two groups. The average error rate is the mean of the error rates across the two groups, and error disparity is the absolute difference in error rates between the two groups. 
The fairness score (FS) is computed as follows:

%\[
%FS = -\alpha \cdot \text{Average Error Rate} - \beta \cdot \text{Error Disparity}; \alpha, \beta >=0
%\]

\begin{equation}
FS = -\alpha \cdot \text{Average Error Rate} - \beta \cdot \text{Error Disparity}, \quad \alpha, \beta \geq 0
\label{eq:fairness_score}
\end{equation}

where:

1. \textbf{Average Error Rate:}

\begin{equation}
\text{Average error rate} = \frac{\text{error}(G_1) + \text{error}(G_2)}{2}
\label{eq:aer}
\end{equation}

%\[
%\text{Average error rate} = \frac{\text{error}(G_1) + \text{error}(G_2)}{2}
%\]

Here, \(\text{error}(G_1)\) and \(\text{error}(G_2)\) are the error rates for the two groups, here the group $G_1$ indicates Normal and $G_2$ refers to CLP.

2. \textbf{Error disparity:}

\begin{equation}
\text{Error Disparity} = \left| \mathrm{error}(G_1) - \mathrm{error}(G_2) \right|
\label{eq:ed}
\end{equation}

%\[
%\text{Error Disparity} = \left| \text{error}(G_1) - \text{error}(G_2) \right|
%\]

3. \(\alpha\) and \(\beta\) are the coefficients to balance the importance of overall error minimization and fairness (disparity minimization).

%The -$\alpha$ and -$\beta$ indicate that higher fairness leads to a higher $FS$.
 The range of $FS$ is $ -\infty\leq \text{FS} \leq 0$. The value of \textbf{\emph{$FS$ closer to zero}} signifies \textbf{\emph{better fairness}} of the system, and vice versa. That is, the value of $FS$ represents the \textbf{\emph{degree of fairness}} of the system.
From the given equation, the average error rate is influenced by error ($G_1$) which is the WER for Normal speech and error ($G_2$) which is WER for CLP speech. If either has a high value, the overall average error rate increases. A lower average error rate indicates a better overall performance. The error disparity is determined by the absolute difference between error ($G_1$) and error ($G_2$). A large disparity indicates a significant difference in ASR performance between normal and CLP speech, meaning the system is less fair. A smaller disparity suggests a more balanced ASR performance between both groups and indicates a more fair system. WER values for normal and CLP speech directly affect both the average error rate and the error disparity, affecting the overall fairness score (FS).

\begin{table}[ht]
\centering
\caption{Fairness of ASR available publicly through Google API, $W_C$, $W_N$, and $FS$ are the obtained WER from CLP and normal test utterances and fairness score, respectively for AIISH \& NMCPC datasets.}
\label{fairness_nmcpc_2}
\begin{tabular}{|c|c|c|c|}
\hline
Dataset & \textbf{$W_N \downarrow$} & \textbf{$W_C \downarrow$} & \textbf{$FS \downarrow$} \\ \hline
AIISH & 93.00    & 98.94 &  -50.95   \\ \hline
NMCPC & 30.21    & 74.27 &  -48.15   \\ 
\hline
\end{tabular}
\end{table}
The fairness score and WER obtained using the Google API are tabulated in Table~\ref{fairness_nmcpc_2}. The scores obtained in the AIISH and NMCPC evaluation sets are $-50.95$ and $-48.15$, respectively. Also, a good system has both a low average error rate and error disparity, resulting in a fairness score close to zero. Fairness scores farther from zero indicate greater unfairness toward CLP speech. The maximum degradation in normal speech is mainly due to the noisy environment. The normal speech is mainly spoken by adult doctors. Contrary to typical ASR benchmarks reporting ~5–10\% WER, the off-the-shelf Google API performed poorly on our pathological-speech test sets. Interestingly, error profiles differed by condition: substitutions dominated for normal speech, whereas insertions dominated for CLP. On the AIISH CLP eval set, Google API returned no transcript for 11 of 209 utterances (~5.3\%); on the NMCPC CLP eval set, it failed on 26 of 210 (~12.4\%). Among the utterances that were transcribed, the average WER was 98.94\% (AIISH) and 74.27\% (NMCPC). These unexpected failure modes motivate the severity-aware modeling and \textcolor{black}{mixing} explored in the next section. We compute the fairness score with 
$\alpha$= $\beta$=0.5; these weights apply only to the fairness score (Eq.~\ref{eq:fairness_score}) and do not affect WER.

%The state of the art WER for ASR is $5-10$\%. It was noted that the main error in the normal case was substitution error whereas for CLP, the main error was insertion error. In AIISH, CLP eval set Google API was not able to provide the text output for $11$ utterances out of $209$ text files, NMCPC CLP eval set, Google API was not able to provide the text output for $26$ utterances out of $210$ text files. The WER of the existing files was obtained as $98.94\%$ for AIISH and $74.27$\% for NMCPC with $\alpha$, $\beta$ equal to $0.5$.

\section{Proposed approach for CLP ASR}

CLP speech exhibits distortions from normal speech, as consistently highlighted in existing literature~\cite{evaluationand, methofperceptualassessment, clpnatureandremediation,  journalofvoice}. These distortions arise due to  physiological differences, leading to unique acoustic and articulatory characteristics in CLP speech~\cite{Zajac2011ReliabilityAV, Whitehill2004SinglewordII}. Figure~\ref{fig2}(a-h), shows the speech signal and the corresponding spectrogram of normal, mild, moderate, and severe speech utterances, respectively. Variations in amplitude and duration suggest differences in speech articulation and phonation among the different severity levels. The visual observation suggests that, compared to normal, the spectral resonance pattern is less distorted in mild cases, and the distortion increases gradually from mild to moderate and severe. To formally assess the observed differences across severity levels, we conducted a distance-based analysis using dynamic time warping (DTW)~\cite{dtw} on three utterances per category (normal, mild, moderate, severe). We first computed spectrograms with a $20 ms$ frame size and a $10 ms$ frame shift. An energy-based speech activity detector (threshold: $6\%$ of the average frame energy) was then applied to retain only speech frames. Using each normal utterance as the reference, we computed DTW distances to the corresponding mild, moderate and severe utterances. Figure~\ref{fig2}(i) presented the distribution of these distances, which increased progressively from normal–mild to normal–moderate and normal–severe. 

%In State of the art API, the fairness is seen to be degrading. To better understand the impact of CLP speech with different degradation to the ASR system, to further investigate the impact of training and testing mismatches, a criss-cross evaluation was performed. 

The fairness of state-of-the-art APIs was observed to degrade in the presence of CLP speech. To gain a deeper understanding of how varying degrees of speech degradation affect ASR performance and to examine the consequences of mismatches between training and testing conditions, a criss-cross evaluation was conducted. In this experiment, models were trained separately on normal and CLP speech and subsequently tested across both types. The aim was to observe how recognition accuracy varied when the speech condition during testing did not align with that used during training. The evaluation encompassed three models GMM-HMM~\cite{GMM}, XLSR~\cite{xlsr}, and WHISPER~\cite{whisper}  and was carried out on two datasets: AIISH and NMCPC~\cite{jawed}.  The GMM-HMM model was chosen as a classical baseline due to its effectiveness in low-resource and well-controlled experimental settings, especially for phoneme-level analysis. It also allows for clear interpretability in terms of acoustic modeling, which is critical when dealing with pathological speech such as CLP. Whisper, a weakly supervised transformer-based ASR model developed by OpenAI, was included to evaluate the performance of modern foundation models trained on large-scale multilingual data. Its robustness and generalization capabilities across diverse acoustic conditions make it a strong candidate for testing fairness and performance on disordered speech. XLSR (Cross-Lingual Speech Representations), a unsupervised model based on the wav2vec 2.0 architecture, was selected for its strong performance in cross-lingual and low-resource settings. Together, these models provide a diverse benchmarking framework that includes both traditional and state-of-the-art systems, enabling a comprehensive evaluation of ASR performance and fairness across varying speech conditions and languages. It is noteworthy that the foundational speech models had minimal exposure to Kannada during pretraining, which likely contributed to their reduced performance on the Kannada-based CLP data. We posit that the English dataset may yield better outcomes, reflecting the pretrained models’ stronger prior exposure to English language and phonetic patterns. Accordingly, the experiments were conducted on both Kannada and English datasets to assess the generalizability of the models across languages with varying levels of representation in pretraining.%In contrast, the English dataset might yield better outcomes, reflecting the foundational model's stronger prior knowledge of the English language and phonetic patterns. Accordingly, the experiments were conducted on both Kannada and English datasets to assess the generalizability of the models across languages with varying levels of representation in pretraining.

The results were evaluated using both WER and Phoneme Error Rate (PER) to assess each model’s ability to generalize across normal and disordered speech. While WER reflects overall transcription accuracy influenced by the language model, PER derived from a monophone-based model offers a more detailed assessment of acoustic performance by capturing phoneme-level misarticulations. %Thus, PER serves as a complementary metric to WER, providing deeper insight into the behavior of the acoustic model, particularly in the presence of speech distortions. 
Insights from the criss-cross analysis, together with Figure 1, indicate that the spectral resonance dynamics of mild CLP are similar to those of normal speech. Motivated by this and prior literature~\cite{dataaugpro}, we hypothesize that adding mild utterances to the CLP training set may improve performance. Accordingly, we perform severity-aware data \textcolor{black}{mixing} to enhance ASR on CLP speech.
%Building on the insights from the criss-cross analysis, and viewing from Figure~\ref{fig2} that spectral resonance dynamics are similar in mild and normal, this motivates that aug of mild with CLP might improve the performance following prior literature~\cite{dataaugpro} and with this hypothesis, data aug is undertaken to improve model performance on CLP speech. %Data augm improves performance by increasing data diversity, reducing the domain gap between normal and CLP speech, and helping models generalize better. 
%The improvement in performance resulting from data aug can be attributed to the similarity in spectral dynamics between normal and mild CLP speech. 

%This similarity allows the model to generalize more effectively, demonstrating the benefit of aug in enhancing recognition accuracy. 
%below in active voice
Finally, this study addresses the issue of fairness in speech recognition by proposed severity-aware \textcolor{black}{mixing} of CLP speech.  
It is hypothesized that a fundamental trade-off exists between maximizing speech recognition performance and ensuring fairness between speakers with CLP and normal speech. By systematically adjusting hyperparameters $\alpha$ and $\beta$ which control the weighting of performance and fairness in model optimization, it will be possible to analyze and achieve an optimal balance between these objectives.
The summary of proposed experiments are illustrated as below:

\begin{itemize} 

    \item A criss-cross evaluation was performed by training and testing models across normal and CLP speech to analyze robustness under mismatched conditions.

    \item The pathology \textcolor{black}{mixing} strategy is intended to enhance model generalization across severity levels and improve ASR performance on disordered speech.

    \item A fundamental trade-off is hypothesized between ASR performance and fairness. This work provides a study with varying the values of  $\alpha$ and $\beta$.

\end{itemize}

\begin{figure}
	\centering
	\includegraphics[height= 150pt,width=300pt]{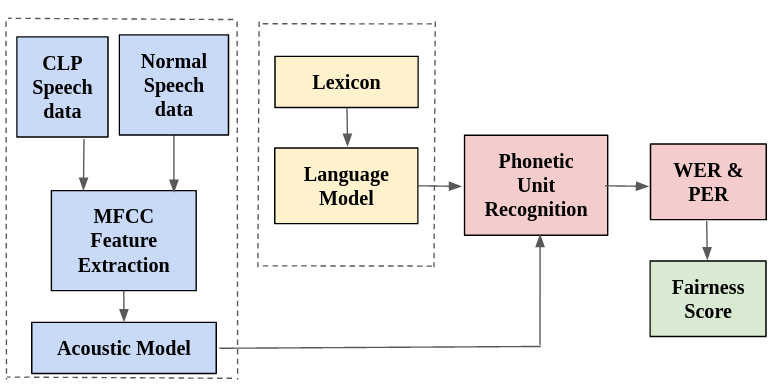}
	\caption{Generalized block diagram of GMM-HMM ASR pipeline.}
	\label{fig3}
 \vspace{-0.5 cm}
\end{figure}

\section{Experimental setup, and results}

\subsection{Experimental setup for ASR models}
%XLS-R and Whisper directly takes input from speech signal and GMM takes MFCC features as input. The Whisper model employs a frame size of $20ms$ with a frame shift of $10ms$, whereas Wav2Vec2 uses a frame size of $25ms$ and a frame shift of 10ms. In the bigram GMM-HMM framework, both triphones have been modeled using $50$ senones and $500$ Gaussian mixtures. The training partition in all the experiments is used to train the wav2vec2 feature extractor for XLS-R and GMM-HMM takes $39$ dimensional MFCC$+\Delta+\Delta\Delta$ feature vectors as input. Wav2Vec 2.0's feature extractor functions as the model's first stage, converting unprocessed audio input into latent speech representations that identify specific waveform patterns. A sequence of CNN layers make up this stage mainly~\cite{wav2vec21, wav2vec22}. The metrics used for estimating the performance are WER and PER. 

%\& \&

XLSR\footnote{https://huggingface.co/facebook/wav2vec2-large-xlsr-53} and Whisper\footnote{\label{hfnote}https://huggingface.co/openai/whisper-small} directly take raw speech signals as input, whereas the GMM-HMM model uses MFCC features. 
\textcolor{black}{Whisper and XLSR-Wav2Vec2 have approximately the same effective window size of $25\,\mathrm{ms}$, but their frame shifts differ: Whisper uses $10\,\mathrm{ms}$, while Wav2Vec2/XLSR uses $20\,\mathrm{ms}$.} In the bigram GMM-HMM framework, triphones are modeled using $50$ senones and $500$ Gaussian mixtures. In all experiments, the training partition is used to train the Wav2Vec2 feature extractor in XLSR. For GMM-HMM, $39$-dimensional MFCC feature vectors with first and second-order derivatives ($\Delta$ and $\Delta\Delta$) are used as input. The feature extractor of Wav2Vec2.0 serves as the first stage of the model, converting raw audio into latent speech representations that capture discriminative waveform patterns. This stage primarily consists of a sequence of convolutional layers~\cite{wav2vec21, wav2vec22}.% developed on Hugging Face which is an open-source machine learning platform\footnote{https://huggingface.co}. 
The performance metrics used for evaluation are Word Error Rate (WER) and Phoneme Error Rate (PER).

\textcolor{black}{The \textcolor{black}{mixing} in training data is done using real CLP speech data in equal proportion which belonged to different severity levels.
Four different training sets were made for the AIISH corpus, by successively increasing the severity classes in order: Only Normal, Mild+Normal, Mild+Moderate+Normal, and Mild+Moderate+Severe+Normal. In all training conditions, the size of the training set was fixed at $304$ utterances, while the number of utterances in the development set was set to $76$. The utterances were equally divided among the various severity groups that formed part of a particular condition as follows: $304/76$ training and development utterances each for normal in only Normal; 152/38 training and development utterances for each of normal and mild in No+Mi; around $101-102/25-26$ training and development utterances each for normal, mild and moderate in No+Mi+Mo; and $76/19$ utterances training and development each for normal, mild, moderate and severe in No+Mi+Mo+Se. Finally, in all AIISH conditions, the external evaluation set was fixed to include $152$ samples, $38$ files from each test set containing normal, mild, moderate and severe data.
In the NMCPC dataset, the same progressive severity-based configuration was followed. There were $280$ training samples and $70$ development samples for each training condition. For the Normal-only condition, there were $280$ training samples and $70$ development samples from the normal population. In the No+Mi condition, the data were balanced by using $140$ training and $35$ development samples for each normal and mild populations. In the No+Mi+Mo condition, the training data was nearly balanced by using $93-94$ training samples for each group and $23-24$ development samples. In the No+Mi+Mo+Se condition, there were $70$ training samples and $17-19$ development samples for each severity level. Finally, in all NMCPC conditions, the external evaluation set was fixed to include $264$ samples, including $66$ samples from each severity level.
The performance metrics include the pooled WER which was calculated to weigh the normal speech and CLP speech equally. The pooled WER, ${\mathrm{WER}_\mathrm{Pooled}}$ is calculated as:}

%\mathrm{WER}_{\mathrm{CLP\text{-}avg}} &=
%\frac{\mathrm{WER}_{\mathrm{Mi}} + \mathrm{WER}_{\mathrm{Mo}} + \mathrm{WER}_{\mathrm{Se}}}{3}, \\
%\mathrm{WER}_{\mathrm{Pooled}} &=
%\frac{\mathrm{WER}_{\mathrm{No}} + \mathrm{WER}_{\mathrm{CLP\text{-}avg}}}{2}.
%\end{align}
%\[
\textcolor{black}{
\begin{align}
\mathrm{WER}_{\mathrm{Pooled}}
=
\frac{\sum S + \sum D + \sum I}{\sum N}
\times 100
\end{align}
%\]
where, ${\sum S}$=Number of substituted words, ${\sum D}$=Number of deleted words, ${\sum I}$=Number of insetrted words, ${\sum N}$=Number of reference words}%$\mathrm{WER}_{\mathrm{No}}$, $\mathrm{WER}_{\mathrm{Mi}}$, $\mathrm{WER}_{\mathrm{Mo}}$ and $\mathrm{WER}_{\mathrm{Se}}$ denote the WER of normal, mild, moderate and severe speech respectively.}

\textcolor{black}{All experiments were conducted using a random seed of $42$. The training and development partitions were first combined, after which stratified 5-fold cross-validation was performed. The folds were stratified by severity category to preserve, as closely as possible, the relative distribution of normal, mild, moderate and severe speech across the training and validation subsets of each fold. The external evaluation data were kept fixed and were not included in fold construction.
For XLSR models, the training was stopped after $120$ epochs or after reaching a minimum improvement of $0.1$ in macro WER on the development split evaluated every three epochs. For the main XLSR experiments, we used the following configuration: learning rate = $(3 \times 10^{-4})$, weight decay = $0.01$, batch size = $16$, eval batch size = $8$, gradient accumulation = $2$, FP16 training, gradient checkpointing and frozen convolutional feature encoder.
For Whisper, the final comparison involved using Whisper-small with full fine-tuning, the maximum number of epochs being $120$, batch size being $16$, evaluation batch size being $8$, gradient accumulation being $4$, FP16 training, gradient checkpointing, and early stopping using development set macro WER. The learning rate used in the experiment was $(5 \times 10^{-6})$. The decoder used was greedy decoder, and the maximum length of the output was $24$ tokens. During training, the balanced-batch mode was enabled to ensure that each mini-batch contained an equal number of samples from the included severity categories. The fold level ensemble decoding was done during the inference stage.}

%; however, the cross-fold parameter averaging was not employed in the main experiments.}
%\textcolor{black}{To provide a fair gender-based WER evaluation when there are different numbers of files in two sets, we have opted for the maximum equal-sized subset selection method. In each case, the size of the subset was determined by the smallest number of files in one of the sets, and all the files belonging to this group were taken into consideration; 57 files were used for AIISH CLP female versus male evaluation since the female CLP set included 57 utterances, and 84 files were used for the earlier NMCPC CLP female versus male evaluation since the female CLP set included 84 utterances. From the bigger male sets, an equal number of files was chosen in order to ensure maximal speaker/utterance diversity.}

\subsubsection{Setup for fairness (\texorpdfstring{$\alpha$}{alpha}, \texorpdfstring{$\beta$}{beta})}

The fairness analysis of the ASR models was conducted using various combinations of $\alpha$ and $\beta$ values from the fairness scoring equation~\ref{eq:fairness_score}, with the constraint $\alpha + \beta = 1$. This allows us to evaluate system performance under different emphases on two metrics: Average Error Rate and Error Disparity. Specifically, we use three combinations of weights: ($\alpha = 0.9$, $\beta = 0.1$), ($\alpha = 0.5$, $\beta = 0.5$), and ($\alpha = 0.1$, $\beta = 0.9$), representing three evaluation perspectives.

%The motivation behind choosing the boundary values ($0.1$, $0.5$, $0.9$) is to analyze how the system behaves when priority shifts between accuracy and fairness. The first case assigns a higher weight to the average error rate ($\alpha = 0.9$, $\beta = 0.1$), which is appropriate when overall accuracy is the priority, even if fairness is compromised.

The first case ($\alpha = 0.9$, $\beta = 0.1$) emphasizes the Average Error Rate, i.e., overall accuracy. This setting is suitable when the application demands high accuracy across all groups, even at the cost of fairness. For instance, if one group significantly outweighs the other in terms of data samples, focusing on minimizing the total error may improve generalization. The second case ($\alpha = 0.1$, $\beta = 0.9$) prioritizes fairness by emphasizing Error Disparity. This is especially relevant in applications like speech disorder classification, where it is essential that underrepresented groups, such as CLP, are not disproportionately misclassified. In such ethically sensitive applications, ensuring equal performance across groups is more critical than maximizing overall accuracy.
The balanced case ($\alpha = 0.5$, $\beta = 0.5$) aims to trade off between accuracy and fairness. This configuration is applicable when both aspects are equally important. %A model might achieve high performance overall but be biased toward one class; the balanced weighting seeks to ensure both effectiveness and equity. 
Depending on the application's requirements, different values of $\alpha$ and $\beta$ may be selected to reflect the desired balance between overall error minimization and fairness across groups.

\textcolor{black}{Based on the analysis of the fairness score, $\alpha$ and $\beta$ play a significant role in interpreting the results obtained from the experiment. In the case where $\alpha$ is high, the ranking is mostly based on the average error rate; hence models having less WER are ranked highly regardless of whether the error distribution is consistent in different severity groups. On the other hand, when the value of $\beta$ is high, the ranking focuses more on the difference among different models, regardless of whether their WER averages are relatively higher than those of other models. It is thus clear that in the fairness-oriented setting, the sensitivity is high concerning poor performance of the ASR system in high or under-represented severity classes. This is critical in ensuring that a lower WER does not equate to fair ASR performance in all types of speech.
For instance, Koenecke et al.~\cite{Koenecke} found that the ASR systems designed for commercial use achieved significantly larger values of WER for Black users in comparison to white users, which means that an average WER can overlook any harm done to particular groups. According to \cite{fairfair} one should consider fairness as a compromise between different criteria since there is no single measure that could be used universally.
This experiment examined the degree of fairness using three different weightings in the fairness scoring metric. The average error rate is denoted as (AER), and the error disparity is represented as (ED).}
%[
%FS_{\alpha,\beta} = \alpha AER + \beta ED, \qquad \alpha+\beta=1.
%]
\textcolor{black}{Here, a lower value of the fairness score means a better trade-off between accuracy and fairness. The variable $\alpha$ determines the importance of overall ASR accuracy, and $\beta$ measures the importance of decreasing disparities among speaker groups. There are three sets of values of $\alpha$ and $\beta$: ($\alpha=0.9,\beta=0.1$), ($\alpha=0.5,\beta=0.5$), and ($\alpha=0.1,\beta=0.9$). The weighting system with parameters ($0.9,0.1$) tends to be accuracy-focused and places relatively high importance on reducing the average error rate. Thus, such a weight setting can be used if the goal is the maximization of overall recognition accuracy. However, in this case, we may overlook the subpar performance of ASR on underrepresented or more clinical groups of speakers. The third setting ($0.1,0.9$) is a fairness-oriented scenario that penalizes heavily any significant error rate disparity among groups. Such an approach is critical in the case of speech disorder ASR, wherein the presence of extremely high error rates on groups of speech pathology may cause uneven accessibility regardless of the general average performance level of the system.
The investigation of these three settings will enable the evaluation of ASR systems in the context of their application. A model that works most effectively with parameters ($\alpha=0.9,\beta=0.1$) should be considered accuracy-oriented, and a model that is most efficient with parameters ($\alpha=0.1,\beta=0.9$) will be considered fairness-oriented. However, if a model ranks well in all three scenarios, it implies that it has low error and is stable in terms of its results for all the speech groups. Otherwise, the variation in model ranking between different values of ($\alpha,\beta$) may imply a certain tradeoff – the reduction of average WER at the expense of increasing disparity, as well as providing stable performance at higher WER levels.
When $\beta$ is increased, there is more weight on the fairness component since the score takes into account Error Disparity more than Average Error Rate. This implies that, when $\beta$ is increased, $\alpha$ is reduced. Hence, when $\alpha$ is lower than $\beta$, the metric is less focused on WER performance and pays more attention to the difference in performance between groups. In this context, an improvement in Average WER does not necessarily translate to an improved fairness score. It might happen that although a model has good performance, it performs better in some groups than others. For instance, it could be the case that the model has extremely low WER on normal data and extremely high WER on severe CLP data. In such a scenario, the model would suffer significantly due to Error Disparity.
Increased $\beta$ indicates fairness-based assessment and increased $\alpha$ denotes accuracy-based assessment. However, one thing must be noted; the trade-off exists only in cases where accuracy and fairness come at odds. In other words, the trade-off arises from situations whereby a model has either low WER with high Error Disparity or high WER with low Error Disparity.}

\subsection{Experimental results}
This section presents the performance analysis of three different ASR models GMM-HMM, XLSR and Whisper on CLP and normal speech. The results are evaluated using WER and PER, and further analyzed for fairness across different severity levels. The initial cross-testing experiments indicate that WER and PER increase as speech severity worsens. When trained and tested on normal speech, the models achieve low WER values, but performance degrades significantly when tested on CLP speech, particularly in the severe category. \textcolor{black}{Mixing training data with real CLP speech from different severity levels improves fairness but does not fully close the performance gap.} The fairness analysis shows that Whisper consistently performs better than XLSR and GMM-HMM, particularly for the NMCPC dataset, whereas GMM-HMM is more suitable for AIISH.% due to its better performance on child speech.

The following subsections provide a detailed breakdown of ASR performance for each model.
%%%%%%%%%%%%%%%%%%%%%%%%%%%%%%%%%%%%%%%%%%%%%%%

\subsubsection{GMM-HMM}

Initially, the ASR is done using GMM-HMM and using the MFCC features trained with NMCPC and AIISH datasets in Kaldi toolkit. %Then in the similar way XLSR and Whisper based transformer models are applied for the same task. 
In a comparable manner, XLSR and Whisper-based transformer models were applied to the same task. The obtained performance of the initial criss cross-experiments is tabulated in Table~\ref{aiish1} and Table~\ref{nmcpc1}. 

\renewcommand{\arraystretch}{1.5}  % Increase row height (default is 1.0)
\begin{table}[]
\centering
\caption{Performance of criss-cross analysis for \textbf{AIISH}. CLP macro: Average CLP test set.}%No, Mi, Mo, and Se represent normal, mild, moderate, and severe respectively. Tot: total CLP test set.}
%\vspace{-0.2 cm}
\fontsize{8 pt}{8 pt}\selectfont
\begin{tabular}{|c|cc|c|cccc|}
\hline
\textbf{Model} &
\multicolumn{2}{|c|}{\textbf{Test$\rightarrow$}} &
  \multirow{2}{*}{\textbf{Normal}} &
  \multicolumn{4}{c|}{\textbf{CLP}} \\ \cline{2-3} \cline{5-8} 
 & \multicolumn{2}{|l|}{\textbf{Train$\downarrow$}} &
   &
  \multicolumn{1}{c|}{\textbf{Mild}} &
  \multicolumn{1}{c|}{\textbf{Moderate}} &
  \multicolumn{1}{c|}{\textbf{Severe}} &
  \textbf{CLP macro} \\ \hline

\multirow{4}{*}{\textbf{GMM-HMM}} &
\multicolumn{1}{|c|}{\multirow{2}{*}{\textbf{Normal}}} &
  WER &
  {3.31} &
  \multicolumn{1}{c|}{27.35} &
  \multicolumn{1}{c|}{32.23} &
  \multicolumn{1}{c|}{87.50} &
  49.02 \\ \cline{3-8} 
 & \multicolumn{1}{|c|}{} &
  PER &
  38.42 &
  \multicolumn{1}{c|}{55.98} &
  \multicolumn{1}{c|}{62} &
  \multicolumn{1}{c|}{76.85} &
  64.94 \\ \cline{2-8}

 & \multicolumn{1}{|c|}{\multirow{2}{*}{\textbf{CLP}}} &
  WER &
  {3.31} &
  \multicolumn{1}{c|}{{9.40}} &
  \multicolumn{1}{c|}{16.53} &
  \multicolumn{1}{c|}{82.50} &
  36.14 \\ \cline{3-8} 
 & \multicolumn{1}{|c|}{} &
  PER &
  52.45 &
  \multicolumn{1}{c|}{53.99} &
  \multicolumn{1}{c|}{59.67} &
  \multicolumn{1}{c|}{75.89} &
  63.18 \\ \hline

\multirow{2}{*}{\textbf{XLSR}} &
\multicolumn{1}{|c|}{\textbf{Normal}} &
  WER &
  {1.65} &
  \multicolumn{1}{c|}{26.50} &
  \multicolumn{1}{c|}{39.67} &
  \multicolumn{1}{c|}{101.67} &
  55.94 \\ \cline{2-8} 
 & \multicolumn{1}{|c|}{\textbf{CLP}} &
  WER &
  {10.74} &
  \multicolumn{1}{c|}{{17.95}} &
  \multicolumn{1}{c|}{25.62} &
  \multicolumn{1}{c|}{103.33} &
  48.96 \\ \hline

\multirow{2}{*}{\textbf{WHISPER}} &
\multicolumn{1}{|c|}{\textbf{Normal}} &
  WER &
  {8.26} &
  \multicolumn{1}{c|}{104.27} &
  \multicolumn{1}{c|}{65.29} &
  \multicolumn{1}{c|}{117.50} &
  95.69 \\ \cline{2-8} 
 & \multicolumn{1}{|c|}{\textbf{CLP}} &
  WER &
  0.83 &
  \multicolumn{1}{c|}{31.62} &
  \multicolumn{1}{c|}{33.88} &
  \multicolumn{1}{c|}{100.83} &
  55.45 \\ \hline

%\multirow{2}{*}{\textbf{WHISPER}} &
%\multicolumn{1}{|c|}{\textbf{Normal}} &
%  WER &
%  {3.31} &
 % \multicolumn{1}{c|}{66.67} &
 % \multicolumn{1}{c|}{52.07} &
%  \multicolumn{1}{c|}{108.33} &
%  75.69 \\ \cline{2-8} 
% & \multicolumn{1}{|c|}{\textbf{CLP}} &
%  WER &
%  2.48 &
%  \multicolumn{1}{c|}{14.53} &
%  \multicolumn{1}{c|}{15.70} &
%  \multicolumn{1}{c|}{61.67} &
%  30.63 \\ \hline

\end{tabular}
\label{aiish1}
\end{table}

\begin{table}[]
\centering
\caption{Performance of criss-cross analysis for \textbf{NMCPC}. CLP macro: Average CLP test set.}%No, Mi, Mo, and Se represent normal, mild, moderate, and severe respectively. Tot: total CLP test set.}
%\vspace{-0.2 cm}
\fontsize{8 pt}{8 pt}\selectfont
\begin{tabular}{|c|cc|c|cccc|}
\hline
\textbf{Model} &
\multicolumn{2}{|c|}{\textbf{Test$\rightarrow$}} &
  \multirow{2}{*}{\textbf{Normal}} &
  \multicolumn{4}{c|}{\textbf{CLP}} \\ \cline{2-3} \cline{5-8} 
 & \multicolumn{2}{|l|}{\textbf{Train$\downarrow$}} &
   &
  \multicolumn{1}{c|}{\textbf{Mild}} &
  \multicolumn{1}{c|}{\textbf{Moderate}} &
  \multicolumn{1}{c|}{\textbf{Severe}} &
  \textbf{CLP macro} \\ \hline

\multirow{4}{*}{\textbf{GMM-HMM}} &
\multicolumn{1}{|c|}{\multirow{2}{*}{\textbf{Normal}}} &
  WER &
  {23.20} &
  \multicolumn{1}{c|}{47.04} &
  \multicolumn{1}{c|}{63.39} &
  \multicolumn{1}{c|}{87.19} &
  65.87 \\ \cline{3-8} 
 & \multicolumn{1}{|c|}{} &
  PER &
  65.74 &
  \multicolumn{1}{c|}{81.37} &
  \multicolumn{1}{c|}{80.71} &
  \multicolumn{1}{c|}{90.21} &
  84.09 \\ \cline{2-8}

 & \multicolumn{1}{|c|}{\multirow{2}{*}{\textbf{CLP}}} &
  WER &
  {65.60} &
  \multicolumn{1}{c|}{49.80} &
  \multicolumn{1}{c|}{34.65} &
  \multicolumn{1}{c|}{74.79} &
  53.08 \\ \cline{3-8} 
 & \multicolumn{1}{|c|}{} &
  PER &
  85.22 &
  \multicolumn{1}{c|}{82.95} &
  \multicolumn{1}{c|}{79.30} &
  \multicolumn{1}{c|}{87.45} &
  83.23 \\ \hline

\multirow{2}{*}{\textbf{XLSR}} &
\multicolumn{1}{|c|}{\textbf{Normal}} &
  WER &
  {42.00} &
  \multicolumn{1}{c|}{59.68} &
  \multicolumn{1}{c|}{76.77} &
  \multicolumn{1}{c|}{91.32} &
  75.92 \\ \cline{2-8} 
 & \multicolumn{1}{|c|}{\textbf{CLP}} &
  WER &
  {80.00} &
  \multicolumn{1}{c|}{79.84} &
  \multicolumn{1}{c|}{87.80} &
  \multicolumn{1}{c|}{92.98} &
  86.87 \\ \hline

\multirow{2}{*}{\textbf{WHISPER}} &
\multicolumn{1}{|c|}{\textbf{Normal}} &
  WER &
  {5.20} &
  \multicolumn{1}{c|}{15.02} &
  \multicolumn{1}{c|}{42.52} &
  \multicolumn{1}{c|}{81.82} &
  46.45 \\ \cline{2-8} 
 & \multicolumn{1}{|c|}{\textbf{CLP}} &
  WER &
  {6.40} &
  \multicolumn{1}{c|}{9.49} &
  \multicolumn{1}{c|}{21.65} &
  \multicolumn{1}{c|}{60.33} &
  30.49 \\ \hline

\end{tabular}
\label{nmcpc1}
\end{table}

The performance in terms of WER and PER shows the least value when training and testing are performed using the normal category. When trained using normal and testing using normal and CLP, the WER of Normal, mild, moderate and severe are $3.31\%$, $27.35\%$, $32.23\%$, and $87.50\%$, respectively for AIISH. Similarly, for NMCPC WER was obtained as $23.20\%$, $47.04\%$, $63.39\%$, $87.19\%$. This shows that the WER increases due to degradation in speech signal introduced by the severity of CLP. Further, by training, the ASR with CLP and testing with normal, mild, moderate, and severe provides the WER of $3.31\%$, $9.40\%$, $16.53\%$, and $82.50\%$, respectively for AIISH and $65.60\%$, $49.80\%$, $34.65\%$, $74.79\%$, for NMCPC.  
%%%%%%
For AIISH data, the performance obtained by training, the ASR with CLP, in terms of WER and PER for GMM-HMM in CLP is $36.14\%$ and $63.18\%$, respectively and $53.08\%$ and $83.23\%$ for NMCPC. 
The experimental results are tabulated in Table~\ref{wer_aiish} and Table~\ref{wer_nmcpc}. From these tables, it can be seen that $FS$ of CLP is better than normal. It indicates that the system trained with only normal data is unfair for CLP compared to the system trained with CLP data which is fair for CLP test set. The result of the system trained with Normal+Mild+Moderate+Severe for AIISH GMM-HMM gives $FS$ score as $-26.95$ indicating this system is better than the other deeplearning methods. For NMCPC, for GMM-HMM, the system trained with Normal+Mild+Moderate gives $FS$ score as $-29.76$.

%The reason this system performs the better than the others is because it
 %Similarly for Whisper, 

\subsubsection{XLSR and Whisper}

Similarly, for XLSR and Whisper, after training with Normal train set, the test performances on CLP WER for NMCPC dataset is $75.92\%$ and $46.45\%$, respectively. The study shows that Whisper model outperforms the best baseline (GoogleAPI at 74.27\% for NMCPC). On the AIISH dataset, which consists of child speech in Kannada, foundation models perform poorly and traditional models achieve better results. In contrast, for the English NMCPC dataset, foundation models perform better. These findings suggest that transformers are particularly effective for low-resource pathological speech, and the improvements in fairness scores further support this claim. 
 %As evidenced by the study, the performance of transformer provides an improvement over the best performance achieved on GoogleAPI which is $74.27\%$ for NMCPC. As AIISH dataset is child speech and in Kannada language so the foundation models were not performing well for it and traditional models are giving better results whereas for NMCPC being in English language, foundation models are giving better results. This justifies the claim that, transformers are best suitable for predicting the speech to text from low resource pathological speech data, the performance of the fairness score also justifies the same.

%%%%%%%%%%%%%%%%%
\begin{table}[]
\centering
\caption{GMM-HMM, XLSR and Whisper WER performance comparison of \textcolor{black}{mixed} data and Fairness ratio \textcolor{black}{with \textbf{\boldmath{$\alpha=0.5$, $\beta=0.5$}}} of \textbf{AIISH} dataset} % No, Mi, Mo, and Se represent normal, mild, moderate, and severe respectively, Tot: total CLP test set.}
\label{wer_aiish}
\resizebox{\textwidth}{!}{%
\begin{tabular}{|c|c|c|c|c|c|c|c|c|c|c|c|c|}
\hline
\multirow{3}{*}{\textbf{Train$\downarrow$}} & \multicolumn{12}{c|}{\textbf{Test$\rightarrow$}} \\ \cline{2-13} 
 & \multicolumn{4}{c|}{\textbf{GMM-HMM}} & \multicolumn{4}{c|}{\textbf{XLSR}} & \multicolumn{4}{c|}{\textbf{Whisper}} \\ \cline{2-13} 
 & \textbf{Normal} & \textbf{CLP} & \textbf{Pooled WER} & \textbf{FS} & \textbf{Normal} & \textbf{CLP} & \textbf{Pooled WER} & \textbf{FS} & \textbf{Normal} & \textbf{CLP} & \textbf{Pooled WER} & \textbf{FS} \\ \hline
\textbf{Normal} & 3.31 & 49.02 & 37.58 & -41.64 & 1.65 & 55.94 & 42.38 & -48.33 & 8.26 & 95.69 & 76.2 & -81.60 \\ \hline
\textbf{CLP} & 3.31 & 36.14 & 27.97 & -30.40 & 10.74 & 48.96 & 38.62 & -38.42 & 0.83 & 55.45 & 46.14 & \textbf{-50.38} \\ \hline
\textbf{Mild+Normal} & 4.13 & 37.85 & 29.44 & -31.58 & 9.92 & 49.30 & 39.25 & -39.31 & 4.13 & 73.87 & 55.32 & -62.53 \\ \hline

\textbf{Mild+Moderate+Normal} & 3.31 & 33.93 & 26.30 & -28.46 & 3.31 & 36.67 & 28.39 & \textbf{-30.87} & 4.13 & 57.67 & 47.39 & \textbf{-50.46} \\ \hline
\textbf{Mild+Moderate+Severe+Normal} & 4.13 & 32.57 & 25.47 & \textbf{-26.95} & 3.31 & 44.52 & 34.24 & -37.72 & 3.31 & 61.27 & 48.02 & -52.99 \\ \hline
\end{tabular}%
}
\end{table}

\begin{table}[]
\centering
\caption{GMM-HMM, XLSR and Whisper WER performance comparison of \textcolor{black}{mixed} data and Fairness ratio \textcolor{black}{with \textbf{\boldmath{$\alpha=0.5$, $\beta=0.5$}}} of \textbf{NMCPC} dataset} %. No, Mi, Mo, and Se represent normal, mild, moderate, and severe respectively, Tot: total CLP test set.}
\label{wer_nmcpc}
\resizebox{\textwidth}{!}{%
\begin{tabular}{|c|c|c|c|c|c|c|c|c|c|c|c|c|}
\hline
\multirow{3}{*}{\textbf{Train$\downarrow$}} & \multicolumn{12}{c|}{\textbf{Test$\rightarrow$}} \\ \cline{2-13} 
 & \multicolumn{4}{c|}{\textbf{GMM-HMM}} & \multicolumn{4}{c|}{\textbf{XLSR}} & \multicolumn{4}{c|}{\textbf{Whisper}} \\ \cline{2-13} 
 & \textbf{Normal} & \textbf{CLP} & \textbf{Pooled WER} & \textbf{FS} & \textbf{Normal} & \textbf{CLP} & \textbf{Pooled WER} & \textbf{FS} & \textbf{Normal} & \textbf{CLP} & \textbf{Pooled WER} & \textbf{FS} \\ \hline
\textbf{Normal} & 23.20 & 65.87 & 54.95 & -48.81 & 42.00 & 75.92 & 66.77 & -50.34 & 5.20 & 46.45 & 35.74 & -38.49 \\ \hline
\textbf{CLP} & 65.6 & 53.08 & 55.96 & -34.24 & 80.00 & 86.87 & 85.39 & -46.13 & 6.4 & 30.49 & 24.12 & -24.10 \\ \hline
\textbf{Mild+Normal} & 24.40 & 48.78 & 42.34 & -33.36 & 47.60 & 67.58 & 61.16 & -40.57 & 4.80 & 35.78 & 27.63 & -29.30 \\ \hline
\textbf{Mild+Moderate+Normal} & 32.00 & 47.89 & 43.64 & \textbf{-29.76} & 48.40 & 67.37 & 61.66 & \textbf{-40.31} & 4.80 & 27.79 & 21.72 & \textbf{-22.35} \\ \hline
\textbf{Mild+Moderate+Severe+Normal} & 41.60 & 53.36 & 50.15 & -30.95 & 68.40 & 77.91 & 75.28 & -42.39 & 6.00 & 29.26 & 23.12 & -23.19 \\ \hline 
\end{tabular}%
}
\end{table}

%Interestingly, though the ASR is trained using CLP, testing with normal does not have much performance degradation. For AIISH and NMCPC datasets respectively, as seen from Table~\ref{wer_aiish}. The WER performance is ($2.39\%$ and $19.03\%$) with the normal training scenario in GMM-HMM. It is $7.53\%$ to $32.93\%$ in XLSR and $2.21\%$ to $7.33\%$ in Whisper for AIISH and NMCPC respectively. This shows that, though the CLP has utterances shifted formant locations, the ASR can model them and provide competitive performance while testing with normal. This may be due to the somewhat intact nature of the formant contour shape in normal and CLP speech. Further, it is also observed that, even with using CLP in training the performance in severe is very high in both GMM-HMM and the foundation models as well. This shows that severe utterances have some random distortion and are difficult to learn through ASR. 
Although the ASR systems were trained on CLP speech, testing on normal speech did not result in significant performance degradation. As shown in Table~\ref{wer_aiish} \textcolor{black}{and Table~\ref{wer_nmcpc}, the WER for} AIISH and NMCPC under normal training with GMM-HMM is $3.31\%$ and $23.20\%$, respectively. For XLSR, the values are $1.65\%$ (AIISH) and $42\%$ (NMCPC), while for Whisper they are $8.26\%$ (AIISH) and $5.20\%$ (NMCPC). These results suggest that, despite the shifted formant locations in CLP speech, ASR models can capture the underlying structure and maintain competitive performance when tested on normal speech. This can be attributed to the relatively preserved contour shapes of the formants across normal and CLP speech. However, when evaluating severe speech, both GMM-HMM and the foundation models perform poorly, indicating that the severe utterances contain random distortions that are particularly difficult for ASR systems to model.
To avoid confusion that the claim of capturing formant contour shape and observed performance is due to the impact of the language model, the performance is evaluated without using LM. The performances in terms of PER also show a similar trend to WER and hence justifies the claim. The nature of ASR to capture the formant contour shape and the stability in performance in the normal test set when trained with CLP, motivates to mix CLP utterances to the training set. This may help in improving the fairness of the system. Motivated by the facts, a \textcolor{black}{mixing} study is performed and obtained performance along with the degree of unfairness or the fairness score tabulated in Table~\ref{alphabeta}.

\begin{figure}
	\centering
	\includegraphics[height= 220pt,width=340pt]{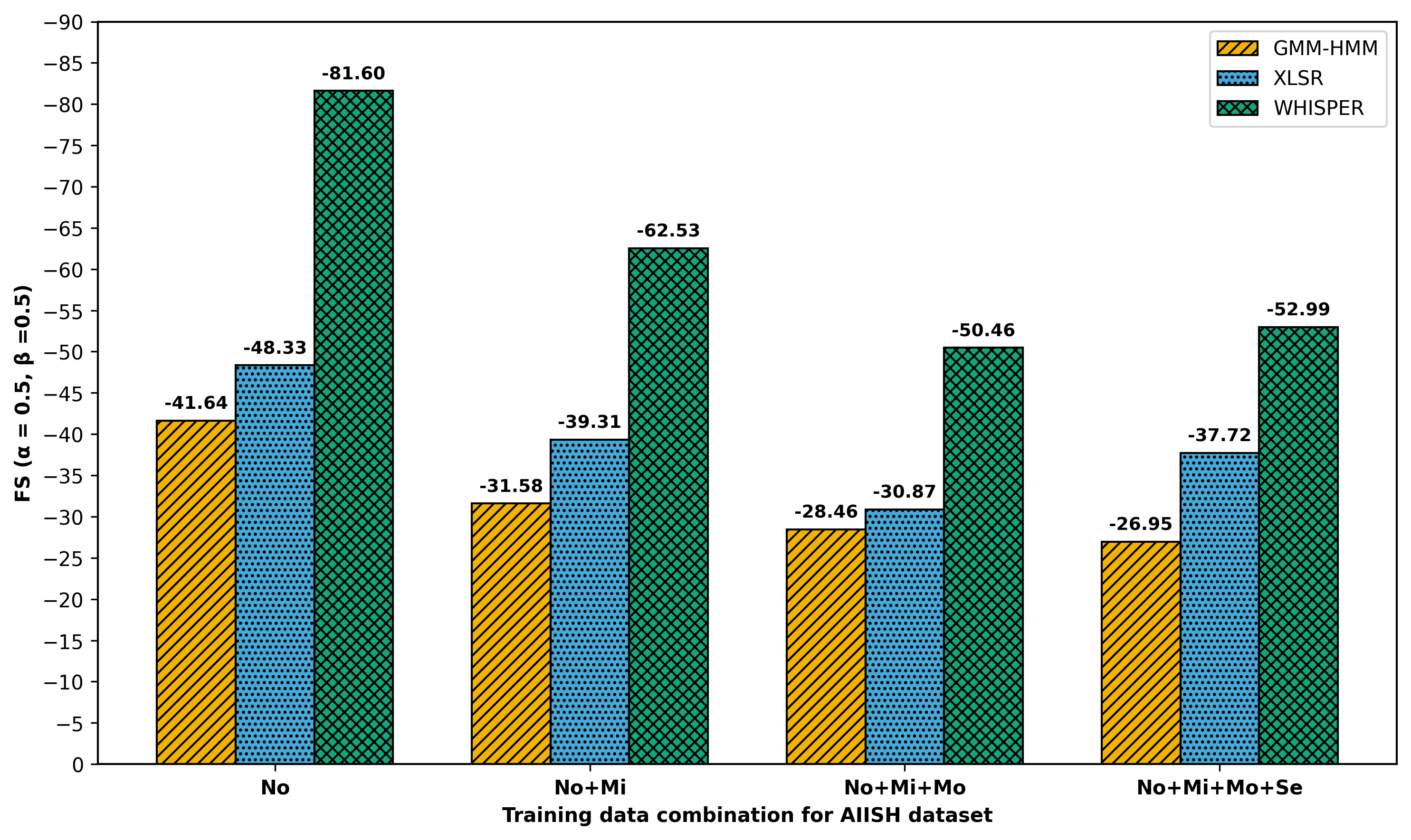}
	\caption{Bar graph showing FS for AIISH dataset. No represents Normal, Mi represents Mild, Mo represents Moderate and Se represents Severe.}
	\label{fig4}

\end{figure}

\begin{figure}
	\centering
	\includegraphics[height= 220pt,width=340pt]{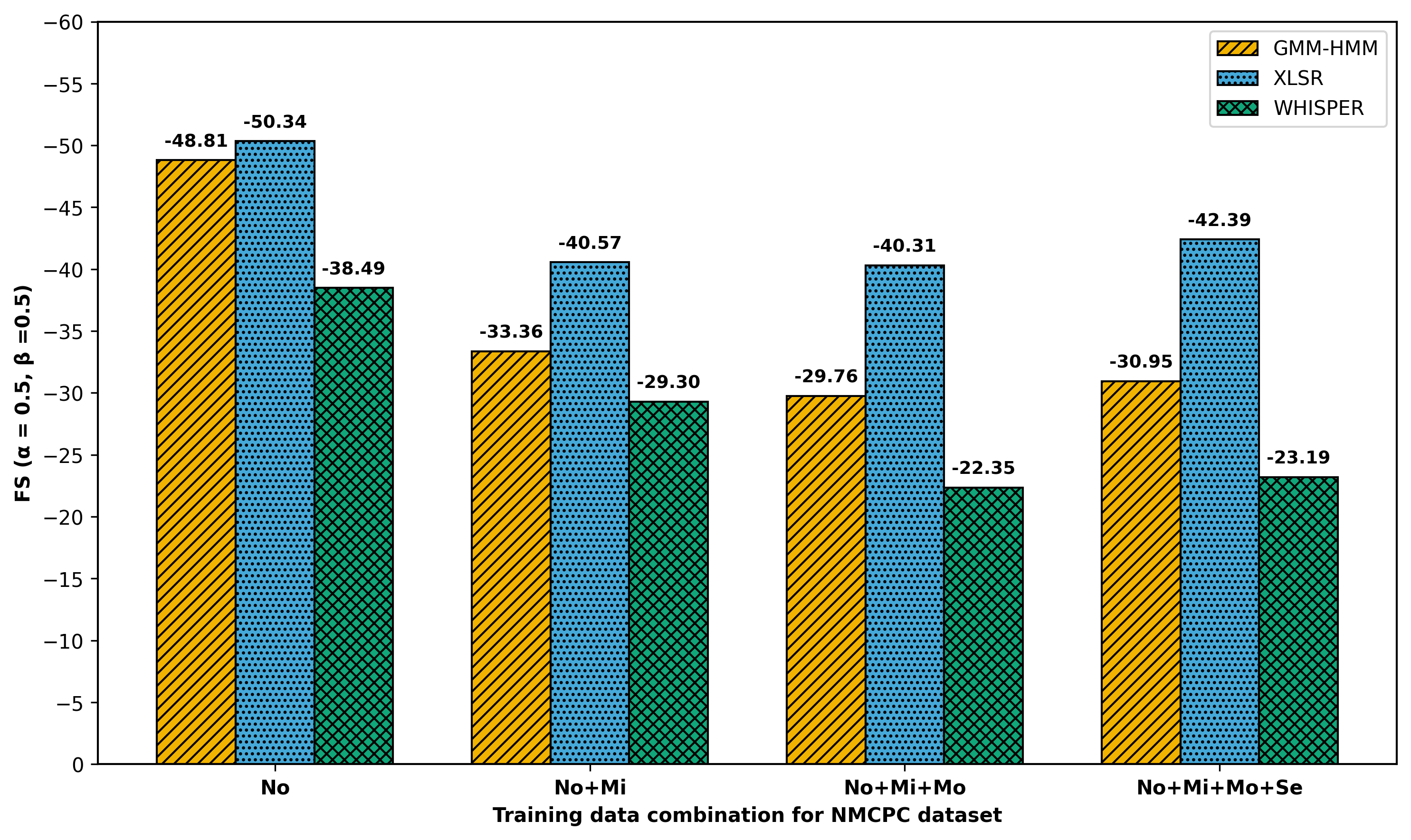}
	\caption{Bar graph showing FS for NMCPC dataset. No represents Normal, Mi represents Mild, Mo represents Moderate and Se represents Severe.}
	\label{fig5}
% \vspace{-0.5 cm}
\end{figure}

\begin{table}[]
\centering
\caption{Alpha and Beta for GMM-HMM \textbf{AIISH} and Whisper \textbf{NMCPC} dataset. }
\label{alphabeta}
\resizebox{\textwidth}{!}{%
\begin{tabular}{|c|c|c|c|c|c|c|c|c|c|c|c|}
\hline
\multirow{3}{*}{\textbf{Train$\downarrow$}} & \multicolumn{10}{c|}{\textbf{Test$\rightarrow$}} \\ \cline{2-11} 
 & \multicolumn{5}{c|}{\textbf{GMM-HMM (AIISH)}} & \multicolumn{5}{c|}{\textbf{Whisper (NMCPC)}} \\ \cline{2-11} 
 & \textbf{CLP} & \textbf{Pooled WER}  & \textbf{\boldmath{$\alpha=0.5$, $\beta=0.5$}} & \textbf{\boldmath{$\alpha=0.1$, $\beta=0.9$}} & \textbf{\boldmath{$\alpha=0.9$, $\beta=0.1$}}

 & \textbf{CLP} & \textbf{Pooled WER}  & \textbf{\boldmath{$\alpha=0.5$, $\beta=0.5$}} & \textbf{\boldmath{$\alpha=0.1$, $\beta=0.9$}} & \textbf{\boldmath{$\alpha=0.9$, $\beta=0.1$}}

 \\ \hline
\textbf{Normal} & 49.02 & 37.58    & -38.39   & -41.64  & -44.90 & 46.45 &  35.74 & -36.29 & -38.49 & -40.70 \\ \hline
\textbf{CLP} & 36.14 & 27.97  & -28.45   & -30.40 & -32.34 & 30.49 & 24.12 & -24.11 & -24.10 & -24.09 \\ \hline
\textbf{Mild+Normal} & 37.85 &  29.44   & -29.86   & -31.58 & -33.29 & 35.78 & 27.63 & -27.96 & -29.30 & -30.65 \\ \hline
\textbf{Mild+Moderate+Normal} & 33.93 & 26.30  & -26.73 & -28.46 & -30.18 & 27.79 & 21.72 & \textbf{-21.84} & \textbf{-22.35} & \textbf{-22.86} \\ \hline
\textbf{Mild+Moderate+Severe+Normal} & 32.57 & 25.47  & \textbf{-25.76}  & \textbf{-26.95}  & \textbf{-28.14} & 29.26 & 23.12 & -23.13 & -23.19 & -23.24 \\ \hline 
\end{tabular}%
}
\end{table}
Based on the data in Table~\ref{wer_aiish}, the fairness score shows a gradual improvement when the system trained on normal speech is incrementally \textcolor{black}{mixed} with speech of increasing severity levels. Specifically, the fairness score improves from $-41.64$ (normal) to $-26.95$ (mild+moderate+severe+normal) in GMM-HMM, from $-48.33$ (normal) to $-30.87$ in XLSR (mild+moderate+normal), and from $-81.60$ (normal) to $-50.46$ (mild+moderate+normal) in Whisper for the AIISH dataset.
%From the tables, it is also observed that when the system is trained using normal, the fairness score improves gradually after aug with severities from mild to mild+moderate and mild+moderate+severe (i.e. increased from $-29.13$ in normal+mild to $-25.92$ in normal+mild+moderate+severe) in GMM-HMM, from $-43.56$ to $-35.88$ in the case of XLSR and from $-37.71$ to $-29.49$ in the case of Whisper respectively for AIISH data as seen from Table~\ref{wer_aiish}. 
Thus the \textcolor{black}{mixing} of the mild, moderate and mild, moderate, severe with the normal training set is able the further improve the fairness of the system. The performance is also evaluated in the NMCPC dataset. Table~\ref{wer_nmcpc} shows the same trend as AIISH - WER decreases after \textcolor{black}{mixing}.
%It is observed from the table that the observed performance shows a similar trend to AIISH. 
Further, the performance of ASR in over CLP test for NMCPC as seen from Table~\ref{wer_nmcpc}, improved by observing the obtained improvement in FS from normal to mild+moderate+normal as $-48.81$ to $-29.76$, in the case of GMM-HMM  from $-50.34$ to $-40.31$ in the case of XLSR and from $-38.49$ to $-22.35$ in the case of Whisper respectively. In most cases, mild+moderate+normal performed better $FS$.
Figure~\ref{fig4} and \ref{fig5} present bar graphs illustrating the FS values for different models (GMM-HMM, XLSR and WHISPER) on the AIISH and NMCPC datasets, respectively. The x-axis represents different subsets of the dataset and refers to speech impairment. The y-axis indicates the FS values, with higher values or in other words, the values closer to zero, generally reflecting better model performance. Across both datasets, the XLSR model (blue) consistently shows the less favorable FS values, whereas GMM-HMM (yellow) and WHISPER (green) exhibit higher values, indicating better performance in AIISH and NMCPC datasets, respectively. WHISPER tends to achieve the lowest FS values in several cases, particularly in the NMCPC dataset, suggesting that it may handle pathological speech more effectively than the other models. The trend in both figures indicates that as the severity of speech impairment increases, the difference in FS values between models becomes more pronounced.
In the Table~\ref{alphabeta}, out of the three cases, one case shows higher weight on average error rate ($\alpha=0.9$, $\beta=0.1$), the second case has a higher weight on error disparity ($\alpha=0.1$, $\beta=0.9$) and finally, the balanced case ($\alpha=0.5$, $\beta=0.5$) is useful when both accuracy and fairness are equally important, ensuring a trade-off between minimizing overall errors and maintaining similar performance across groups. The bold values show which \textcolor{black}{mixing} is favorable for both cases. For NMCPC, Mild+Moderate+Normal system provides better WER as compared to other systems. But for the AIISH dataset, Mild+Moderate+Severe+Normal system gives the best WER for GMM-HMM system which performed better than the CLP system as well. Depending on the user's specifications, different operating systems can be designed accordingly as per the best performance observation of FS.

%This shows the aug of CLP data helps in improving the normal speech ASR performance as well along with the CLP speech. 
Overall, we show that \textcolor{black}{mixing} utterances across severity levels yields measurable gains in ASR accuracy and fairness.
%Finally, the study concludes the fairness and the performance of the ASR system can be improved by aug the utterances from different severity levels during ASR training. 
The fairness score across all the results depict that there is still a wide gap in speech recognition of CLP speech data. The criss cross results have bridged the gap a little and a detailed results shown in the experiments show that \textcolor{black}{mixing} would help further in the improvement of fairness score.

\subsubsection{Study on the Impact of Gender on the Fairness Score}

Since the datasets are imbalanced according to gender, this study was conducted to check the impact of this gender imbalance on FS. 
Table~\ref{tab:gender_wise_wer_whisper} illustrates the WER and Fairness Scores (FS) for Whisper under the different training mixtures in the NMCPC CLP evaluation dataset for male and female speakers. In the case of both male and female speakers, the WER of the CLP speech is always significantly higher than that of the normal speech independently of the training mixture. For instance, in NoMiMo training mixture, the WER for the normal speech of female speakers is 10.00\%, whereas the WER for CLP speech is 36.16\%. Likewise, for male speakers, WER is 4.09\% for normal speech and 23.03\% for CLP speech. It clearly shows that the pathology of the speech has much more effect on the recognition than the speaker gender, as the pathology affects articulation, speech intelligibility and acoustic variability. Moreover, the dataset contains more male speakers than female speakers. So, consequently the WER is higher for females as compared to males.

Incorporating the CLP speech in the training data considerably improves the pooled WER and the fairness score for the male and female speakers in all the training mixtures. NoMiMo training mixture performs the best as it results in the best pooled WER along with the FS value near to zero for both male and female. Incorporation of the severe pathology (NoMiMoSe) leads to insignificant improvements, sometimes even a degradation of both pooled WER and fairness.
The consistently higher WER recorded for the pathological CLP speech when compared to normal speech regardless of gender suggests that the primary reason for reduced recognition efficiency is the presence of pathological speech, not the difference in gender. Pathological speech changes articulation, phoneme production, timing and acoustics and makes recognition more difficult for the ASR system.
Though gender-wise performance results are given, in the normal evaluation set there are only 8 utterances from females compared to 58 utterances from males. Therefore, the difference in ASR performance related to gender needs to be taken with caution due to the limited amount of samples from females.
Further study of gender fairness should be conducted with the use of the larger corpus with an equal amount of samples from males and females of different levels of pathology.

\begin{table*}[htbp]
\centering
\caption{Gender-wise WER on the NMCPC CLP evaluation set for Whisper.
No: Normal, NoMi: Normal+Mild, NoMiMo: Normal+Mild+Moderate, and
NoMiMoSe: Normal+Mild+Moderate+Severe.}
\label{tab:gender_wise_wer_whisper}

\renewcommand{\arraystretch}{1.20}
\setlength{\tabcolsep}{5pt}

\resizebox{0.75\textwidth}{!}{%
\begin{tabular}{|c|c|c|c|c|c|c|c|}
\hline

\multirow{2}{*}{\textbf{Trained Model}} &
\multirow{2}{*}{\textbf{Gender}} &
\multicolumn{6}{|c|}{\textbf{Whisper}} \\
\cline{3-8}

&
&
\textbf{Normal} &
\textbf{CLP} &
\textbf{Pooled WER} &
\textbf{$\alpha$=0.5,$\beta$=0.5} &
\textbf{$\alpha$=0.1,$\beta$=0.9} &
\textbf{$\alpha$=0.9,$\beta$=0.1} \\
\hline

\multirow{2}{*}{\textbf{No}}
& \textbf{F}
& 10.00
& 62.89
& 58.33
& -55.61
& -53.43
& -57.78 \\
\cline{2-8}

& \textbf{M}
& 4.09
& 29.02
& 18.80
& -21.86
& -24.31
& -19.41 \\
\hline

\multirow{2}{*}{\textbf{CLP}}
& \textbf{F}
& 10.00
& 37.11
& 34.77
& -30.94
& -27.87
& -34.00 \\
\cline{2-8}

& \textbf{M}
& 5.45
& 26.50
& 17.87
& -19.46
& -20.73
& -18.18 \\
\hline

\multirow{2}{*}{\textbf{NoMi}}
& \textbf{F}
& 10.00
& 48.43
& 45.11
& -41.77
& -39.09
& -44.44 \\
\cline{2-8}

& \textbf{M}
& 3.64
& 25.24
& 16.38
& -18.99
& -21.07
& -16.90 \\
\hline

\multirow{2}{*}{\textbf{NoMiMo}}
& \textbf{F}
& 10.00
& 36.16
& 33.90
& -30.03
& -26.93
& -33.12 \\
\cline{2-8}

& \textbf{M}
& 4.09
& 23.03
& 15.27
& -17.10
& -18.57
& -15.63 \\
\hline

\multirow{2}{*}{\textbf{NoMiMoSe}}
& \textbf{F}
& 10.00
& 35.85
& 33.62
& -29.73
& -26.62
& -32.84 \\
\cline{2-8}

& \textbf{M}
& 5.00
& 23.66
& 16.01
& -17.33
& -18.39
& -16.27 \\
\hline

\end{tabular}%
}

\vspace{1mm}

\begin{minipage}{0.75\textwidth}
\footnotesize
\textbf{Note:} The NMCPC normal evaluation set contains 8 female and 58 male utterances. The NMCPC CLP evaluation set contains 84 utterances for each gender.
\end{minipage}

\end{table*}

\subsubsection{Effect of $\alpha$ and $\beta$ on mixing configuration}
%Figure~\ref{fig2}

\begin{figure}
	\centering
	\includegraphics[height= 200pt,width=350pt]{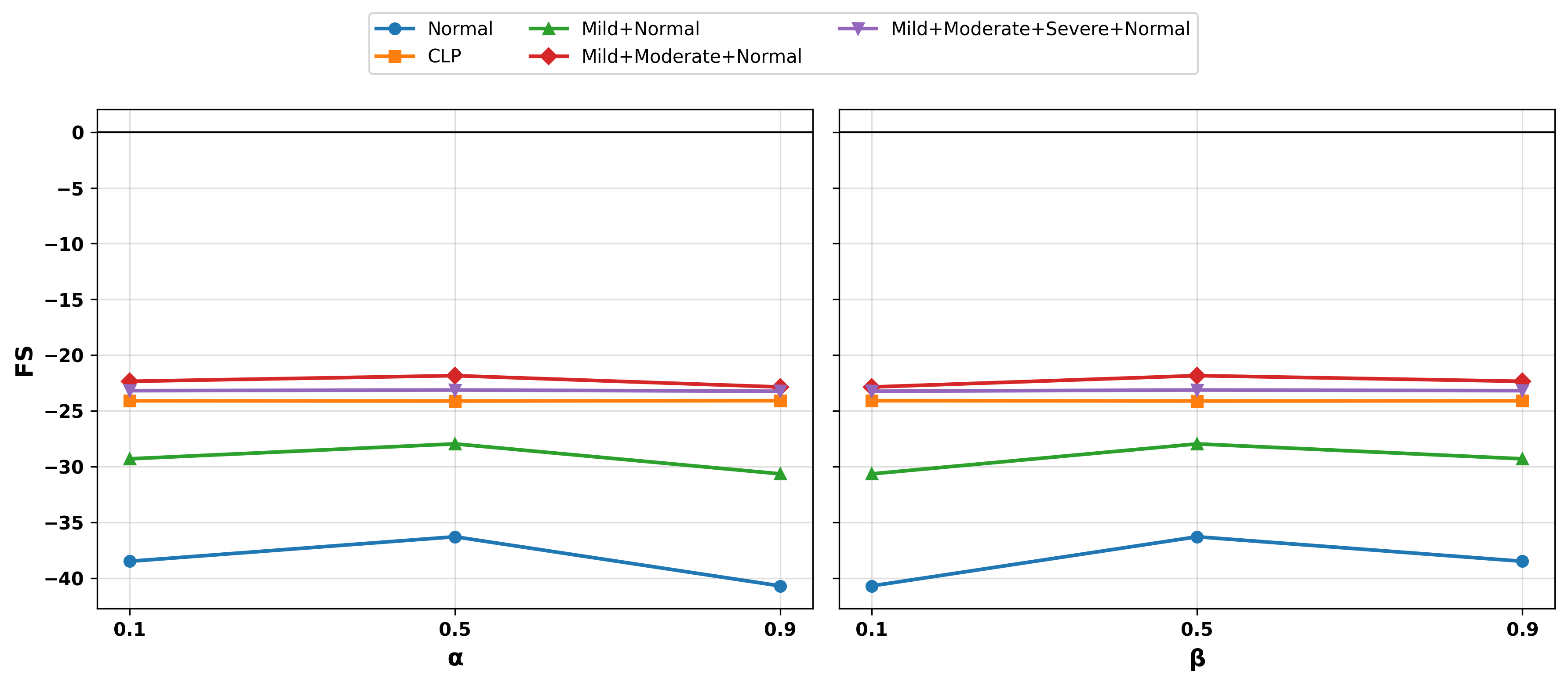}
	\caption{Fairness Vs $\alpha$ and $\beta$ of NMCPC dataset}
	\label{beta}
 \vspace{-0.5 cm}
\end{figure}

By changing $\alpha$ and $\beta$ value it is observed if mixing gives better performance and reduction in disparity between groups. %So including $\beta$ is not required and we can see the same from graphs compared to normal. We can see from the graph that it is flat in nature. 
Two plots in figure~\ref{beta} show how the fairness score is affected by the change in the weight of the importance of the average error and error disparity. In this context, $\alpha$ denotes the weight of the average error rate for both the Normal and CLP groups, while $\beta$ denotes the weight of the absolute performance gap between them.
It depicts how changing the weighting parameter of $\beta$ affects the value of the fairness score for different configurations of training. The balanced weighting ($\alpha$=$\beta$=0.5) always results in the maximum fairness score (the value closest to zero), showing the smallest difference between the two demographic groups. %Changes in the value of the fairness score when moving away from the balanced weighting thus prove that too much attention paid to either group leads to the creation of performance differences. 
In addition, the models with a larger variety of pathological speech in their training set have better performance and fairness score with reduced disparity between groups.
For most Whisper training conditions, the optimal condition, which yields the fairness score closest to zero, is the one where $\alpha=\beta=0.5$. This means that taking into account the overall recognition rate and the absolute performance gap gives a more favorable result in terms of fairness score.% than when any single parameter is overly prioritized.
However, to analyse the disparity between groups, higher $\beta$ value gives a better understanding. Normal only condition demonstrates the greatest variations for different $\alpha$ and $\beta$ values, while other conditions demonstrate greater stability.
As the main goal of the study in terms of fairness is to investigate the difference in the performance between typical and atypical speech, the difference factor $\beta$ is emphasised. The larger the value of $\beta$, the more weight the WER difference between the two groups will have in the score. Thus, the higher the value of $\beta$, the more sensitive the score becomes to the differences between the groups.

\subsubsection{Accuracy-fairness Trade-off}
\begin{figure}
	\centering
	\includegraphics[height= 220pt,width=350pt]{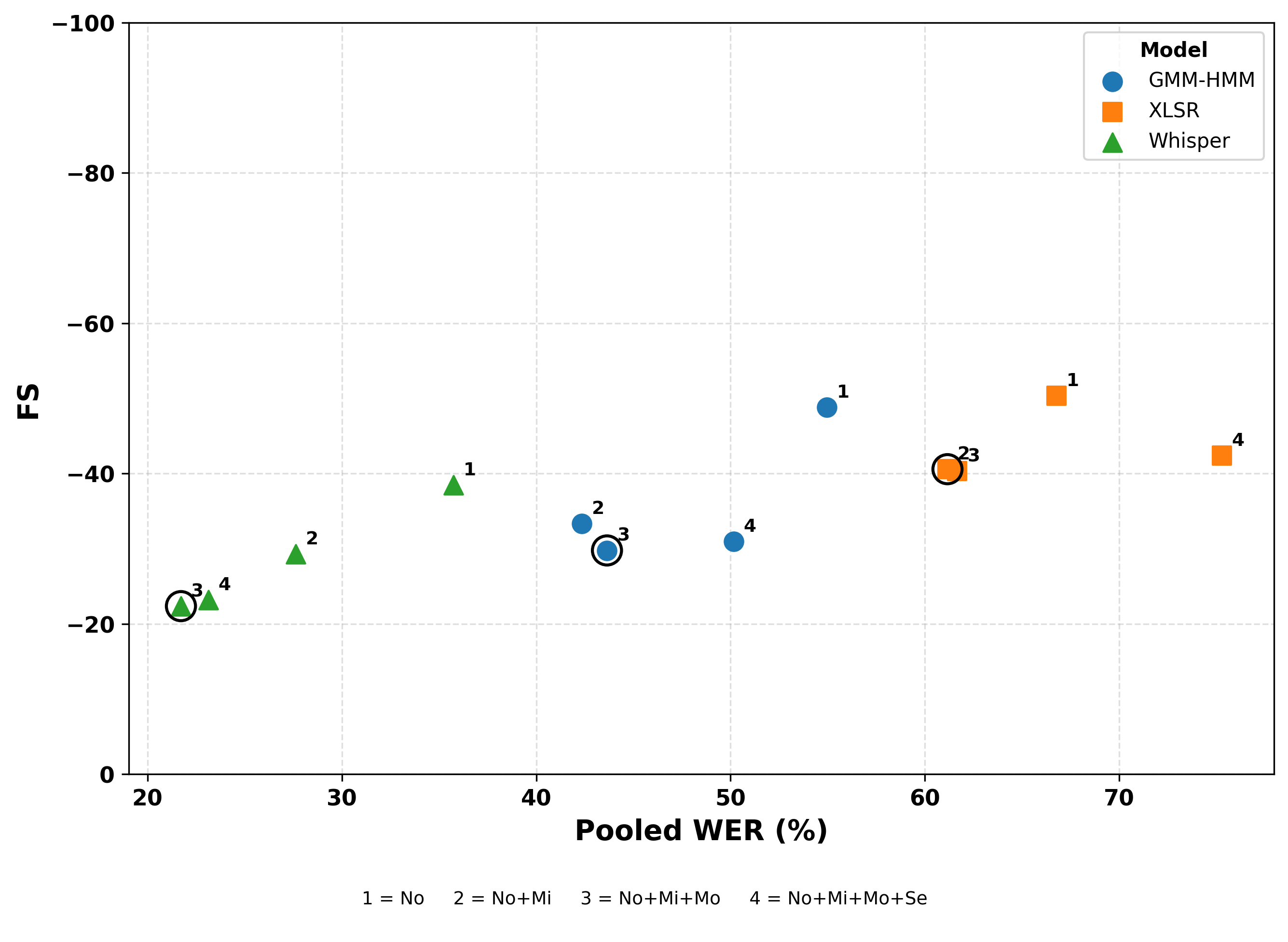}
	\caption{Accuracy-fairness trade-off of NMCPC dataset}
	\label{tradeoff}
 \vspace{-0.5 cm}
\end{figure}

The figure~\ref{tradeoff} examines both the performance of ASR (Pooled WER) and the FS for $\alpha=0.5$ and $\beta=0.5$ at the same time. 
The x-axis shows the Pooled WER (\%) values, and lower values mean good ASR performance. The y-axis shows the Fairness Score (FS), and the lower FS value indicates higher fairness of pathological speech. Thus, the bottom-left of the chart contains the points for models with better ASR performance and fairness. Outlined points indicate equal-weight accuracy-fairness compromise choices.
The points are numbered depending on the training mixtures (1=No, 2=No+Mi, 3=No+Mi+Mo, 4=No+Mi+Mo+Se). From table~\ref{wer_nmcpc}, it is seen that Pooled WER decreases from $35.74\%$ to $21.72\%$ ($\approx 14$ percent improvement). FS changes from $-38.49$ to $-22.35$, which shows that there is a considerably smaller difference between normal and pathological speech. In other words, training mixture 3 (No+Mi+Mo) approaches the desired bottom-left point and is the most optimal compromise between accuracy and fairness for Whisper. The addition of severe speech (training mixture 4) leads to some deterioration (WER increases from $21.72\%$ to $23.12\%$, and FS changes from $-22.35$ to $-23.19$), which means that adding severe pathological speech does not always provide enhancement.
From the three ASR models used for the NMCPC dataset, it can be seen that Whisper is always in the leftmost part of the graph, meaning its pooled WER is much lower than GMM-HMM and XLSR.

\section{Discussion}
Initially, a criss-cross evaluation was conducted using the GMM-HMM system on the AIISH dataset to examine the impact of mismatched training and testing conditions. In this setup, the ASR system was trained separately on normal and CLP speech, and tested on both categories across varying severities (normal, mild, moderate, severe). The primary objective was to analyze the recognition performance under different configurations, particularly to understand how a system trained on CLP speech performs when tested with normal utterances, and vice versa. This evaluation provided critical insights into the generalizability and robustness of the ASR system, revealing significant performance degradation when there was a mismatch between training and testing speech conditions.

Based on these findings, a series of data \textcolor{black}{mixing} strategies were designed to mitigate the performance gap. The underlying hypothesis was that the degree of acoustic distortion increases from mild to severe CLP cases, and therefore, mild CLP speech, being relatively less distorted, could serve as an effective \textcolor{black}{mixing} source when combined with normal speech. This would ideally improve recognition performance on CLP speech without compromising the accuracy on normal speech. Accordingly, multiple GMM-HMM systems were trained with progressively \textcolor{black}{mixed} datasets: (1) normal + mild, (2) normal + moderate, (3) normal + severe, (4) normal + mild + moderate, and (5) normal + mild + moderate + severe.

Subsequently, each of these trained models was evaluated using test sets comprising normal, mild, moderate and severe utterances. A fairness analysis was performed by computing the Fairness Score ($FS$), which quantifies the performance disparity between normal and CLP speech. The goal was to identify the \textcolor{black}{mixed} configuration that yields the most balanced performance, maximizing overall recognition accuracy while minimizing bias across speaker groups. Finally, to validate the consistency of the observations, the entire set of experiments was replicated on the NMCPC dataset, which contains English utterances from a different demographic and linguistic background.

\section{Conclusion}
\label{sec:5}
%It also shows us in the spectro-temporal representation of speech, the shape of the resonance contour is almost intact in normal and mild and degrades gradually from moderate to severe. 

This study presents a comprehensive fairness-centric evaluation of ASR systems for individuals with CLP speech. A criss-cross experimental framework was initially employed to analyze the performance disparity when ASR models trained on normal and CLP speech were tested across varying CLP severity levels. The findings revealed a marked degradation in recognition accuracy, particularly for moderate and severe CLP speech, highlighting the need for inclusive training strategies.

To address this, a series of data \textcolor{black}{mixed} experiments were conducted by combining normal speech with CLP utterances of different severities. The results demonstrated that stepwise severity \textcolor{black}{mixed} with CLP speech alone substantially improved recognition performance across all severity levels, with minimal adverse impact on normal speech recognition. This supports the hypothesis that mildly impaired speech, being acoustically closer to normal speech, serves as an effective \textcolor{black}{mixed} source for enhancing both accuracy and fairness.

Furthermore, transformer-based cross-lingual ASR models, such as XLSR and Whisper, consistently outperformed traditional GMM-HMM models implemented using Kaldi. \textcolor{black}{The handcrafted features in GMM-HMM might also be a reason for the relatively low value for fairness.} These models achieved lower Word Error Rates (WER), especially in English-language CLP speech, indicating their superior generalisation capabilities in low-resource pathological settings. 
\textcolor{black}{Training the ASR model from scratch is out of scope for this research because of the insufficient amount of pathological speech data and the computational cost. In the future, more CLP data can be gathered to examine the effectiveness of training from scratch in comparison with fine-tuning alongwith mitigating gender skewness.}
Overall, it was observed that recognition performance was highest for mild CLP cases and lowest for severe cases, confirming the gradient of intelligibility based on severity. The study underscores the importance of considering speech diversity in ASR training and validates that multi-severity \textcolor{black}{mixed} can lead to fairer and more robust ASR systems. Future work will focus on designing severity-invariant frameworks that can maintain high performance regardless of the degree of speech impairment in CLP patients.

%The work proposes a fairness metric to calculate the fairness of CLP speech. The criss-cross experiment shows that though the training done in CLP by aug mild, moderate, and severe cases improves the performance of inferencing. Training with several sets of CLP severity helps to further draw a conclusion about how the ASR system will perform if aug with pathological speech. XLS-R and Whisper-based cross-lingual ASR performs better as compared to GMM-HMM based KALDI models with a decrement in WER. This shows that transformer based models outperform for ASR tasks for low resource pathological speech in English language. It has also been observed that the speech recognition rate is highest for mild CLP and least for severe CLP. Therefore, our future attempts will try to develop a better framework, that can provide enhanced speech recognition system whose performance should be independent less affected by the variations of CLP patient's severity. The aug study confirms that in the future, the fairness of the ASR system can be improved by aug the utterances from different severity levels during ASR training.

%%%%%%%%%%%%%%%%%%%%%%%%
%%%%%%%% 5th June%%%%%%
% \begin{acknowledgments}
% This research was supported by  ...
% \end{acknowledgments}

\appendix

%\section{Data generation: Language and speaker change detection by humans}\label{SUB_S_DG}

%The dataset used for this study consists of $32$ utterances having language change and $15$ utterances having speakers change. The selected utterances are split around the change point to generate the monolingual/speaker utterances in each case. After splitting, $32$ and $15$  mono-lingual and mono-speaker utterances, that are having higher duration are considered. Actual change points are known as true change points for two-language/speaker utterances, while the beginning of the middle voiced frame is known as the false change point for mono-language/speaker utterances.

\bibliographystyle{elsarticle-num-names} 
\bibliography{sampbib}
\end{document}